\documentclass[preprint2]{aastex}

\usepackage{graphics,graphicx}
\usepackage{natbib}
\usepackage{amsmath, amssymb}
\usepackage[version=3]{mhchem}
\bibpunct{(}{)}{;}{a}{}{,}

\newcommand{\nspe}[1]{n(\ce{#1})}

\newcommand{\kref}[1]{k_\text{\ref{#1}}}
\newcommand{\sigref}[1]{\sigma_\text{\ref{#1}}}
\newcommand{\ki}{\kref{eqOHf1}}
\newcommand{\kii}{\kref{eqH2Of1}}
\newcommand{\kiii}{\kref{eqH2Ophd1}}
\newcommand{\kiiiz}{\kref{eqH2Ophd1}{}_{,0}}
\newcommand{\kiv}{\kref{eqOHphd1}}
\newcommand{\kivz}{\kref{eqOHphd1}{}_{,0}}
\newcommand{\sigiii}{\sigref{eqH2Ophd1}}
\newcommand{\sigiv}{\sigref{eqOHphd1}}

\newcommand{\ddz}{\frac{\mathrm{d}}{\mathrm{d}z}}
\newcommand{\refeq}[1]{equation~(\ref{#1})}
\newcommand{\reffig}[1]{Figure~\ref{#1}}

\newcommand{\FLya}{F_{\text{Ly}\alpha}}

\newcommand{\reftab}[1]{Table~\ref{#1}}

\begin{document}

\title{Water vapor distribution in protoplanetary disks}

\author{Fujun~Du and Edwin~A.~Bergin}
\affil{Department of Astronomy,
       University of Michigan,
       500 Church Street,
       Ann Arbor, MI 48109, USA}
\email{fdu@umich.edu}

\shorttitle{Water vapor in protoplanetary disks}
\shortauthors{Du \& Bergin}

\begin{abstract}
Water vapor has been detected in protoplanetary disks.  In this work we model
the distribution of water vapor in protoplanetary disks with a thermo-chemical
code.  For a set of parameterized disk models, we calculate the distribution of
dust temperature and radiation field of the disk with a Monte Carlo method, and
then solve the gas temperature distribution and chemical composition.  The
radiative transfer includes detailed treatment of scattering by atomic hydrogen
and absorption by water of Ly~$\alpha$ photons, since the Ly~$\alpha$
line dominates the UV spectrum of accreting young stars.  In a fiducial model,
we find that warm water vapor with temperature around 300~K is mainly
distributed in a small and well-confined region in the inner disk.  The inner
boundary of the warm water region is where the shielding of UV field due to
dust and water itself become significant.  The outer boundary is where the dust
temperature drops below the water condensation temperature.  A more luminous
central star leads to a more extended distribution of warm water vapor, while
dust growth and settling tends to reduce the amount of warm water vapor.  Based
on typical assumptions regarding the elemental oxygen abundance and the water
chemistry, the column density of warm water vapor can be as high as
$10^{22}$~cm$^{-2}$.  A small amount of hot water vapor with temperature higher
than $\sim$300~K exists in a more extended region in the upper atmosphere of
the disk.  Cold water vapor with temperature lower than 100~K is distributed
over the entire disk, produced by photodesorption of the water ice.
\end{abstract}


\section{Introduction}

Low mass and possibly high mass stars gain additional mass through a
circumstellar disk at their late stage of formation.
As the disk itself evolves planets are born in the dense dusty midplane,
hence these systems are called protoplanetary disks.  The
physical and chemical environments of the disk is thus vital for determining
the properties of these planets.  Among all the chemical species in a disk,
water is one of the most important, because: (1) it
may carry most of the oxygen that is available, the only competitors being
CO and possibly \ce{CO2} \citep{Favre2013,Pontoppidan2014}; (2) it may contribute
significantly to the heating and cooling of the disk material, hence affecting
the dynamics; (3) it may shield the disk material from UV radiation
\citep{Bethell2009}; (4) it may aid in the coalescence of dust particles to
form planetesimals \citep{Stevenson1988,Ros2013}; (5) its relation to the
origin and sustaining of life.

Water vapor has been detected in protoplanetary disks through infrared
rovibrational and rotational lines \citep{Carr2004, Carr2008, Salyk2008,
Salyk2011, Pontoppidan2010b, Pontoppidan2010a, Doppmann2011, Hogerheijde2011,
Riviere2012, Fedele2012, Najita2013}.
For the most part these observations are spatially and spectrally unresolved,
leaving some uncertainty regarding the overall spatial distribution of water
vapor within the disk.    However, the combination of \textit{Spitzer} and
\textit{Herschel} data provide access to transitions arising from a wide range
of energy states (ground, 0~K, to thousands of K); in this case the abundance
distribution of water vapor might be inferred using other information to
constrain the physical structure (e.g. density and temperature).
In one intriguing study,
\citet{Zhang2013} infer a warm, narrow, and high column density ring of water
at a distance of 4~AU to the central star in TW~Hya by fitting to the \textit{Spitzer}
and \textit{Herschel} spectra.  The water vapor temperatures assumed by these observers
for fitting their data apparently show a dichotomy.  The cold water
vapor has temperatures $\lesssim$100~K, and the hot/warm water vapor has
temperatures of 200--1500~K.  Though never directly spatially resolved, model
fittings in these works suggest that the hot water are concentrated in a small
region close to the central star, and the cold water are distributed over an
extended region in the outer disk.


A few questions naturally arise. (1) How are the water molecules formed in
these disks?  (2) What is the interstellar
heritage of water in the disk?  Did all water form in the prestellar core prior to stellar birth?  (3) What environmental factors determine the presence of water, and
which region and which evolution stage of the disk does the observed water
trace?  (4) Where do these water molecules ultimately go?  (5) Are they related
to the water found on planets and comets, and if related, how?  The present
work will not be able to answer all these questions, but will only contribute
to the understanding of questions 1 and 3.  Some recent studies related to
question 2 and 5 can be found in \citet{Furuya2013}, \citet{Cleeves2014},
and \citet{Albertsson2014}; see also the review by \citet{vanDishoeck2014}.

Among the many modeling efforts devoted to the chemistry of protoplanetary
disks (for recent reviews, see \citealt{Henning2013} and \citealt{Dutrey2014}),
a few have focused on gaseous water.  In \citet{Glassgold2009} high water
abundances is obtained in the molecular transition layer of the inner disk
heated by X-rays.  They emphasize the role of \ce{H2} formation, since \ce{H2}
is a precursor of water.  \citet{Bethell2009} point out that when dust is
settled, water becomes the dominant absorber of UV photons, shielding itself
from UV dissociation.  This self-shielding effect also limits the column
density of warm water vapor to ${\sim}10^{18}$~cm$^{-2}$ and that of OH to ${\sim}2{\times}10^{17}$~cm$^{-2}$.  The model of
\citet{Woitke2009a} shows that in Herbig Ae protoplanetary disks water is
distributed in three regions with distinct properties: a deep warm region, an
irradiated hot region, and a photodesorbed cold region.  \citet{Kamp2013}
caution that the interpretation of the observed water emission is affected by
uncertainties in the chemical input data and radiative transfer.  The recent
work of \citet{Adamkovics2014} focus on the role played by the
photodissociation of water and OH.  In their model warm water is limited to the
inner 4~AU of the disk.

In this work we follow a parameterization approach to study the chemistry of
warm water vapor.  Our goal is to identify the main stellar and disk parameters
that determine the water repository, specifically we clarify the role of water
self-shielding in maintaining its abundance, and in this paper we are \emph{not} aiming to
reproduce any specific observational results, which will be the content of a
follow-up work.  Section~\ref{secModel} contains description of
a new code we have created from scratch for this study.  In
section~\ref{secResults} we present the results, and we conclude our paper in
section~\ref{secConclusion}.

\section{Details of the Modeling}
\label{secModel}

\subsection{Code Description}
\label{secCode}

The layout of our code\footnote{Our code is publicly available at\\
\url{https://github.com/fjdu/rac-2d}} is similar to the
P\textsc{ro}D\textsc{i}M\textsc{o} code \citep{Woitke2009}.  Given a
distribution of gas and dust, we first solve the dust temperature distribution
with a Monte Carlo method based on the strategy of \citet{Bjorkman2001} (see
also \citealt{Baes2005, Bruderer2012}), in which the dust temperature of a
spatial cell is updated each time a photon packet cross this cell; this
strategy is also adopted in the RADMC code \citep{Dullemond2004}.  Although the
geometry in our code is symmetric with respect to rotation around the central
axis and reflection about the midplane, the photon propagation is done in full
three dimensions.  We have not yet implemented the diffusion approximation \citep{Min2009}
for the highly shadowed region where the photon statistics is low, and we
mitigate this by allowing a large number (${\sim}10^7$) of photon packets in
the Monte Carlo.  The whole spectrum (from UV to sub-millimeter) of the central star is
used as input.  Observation and modeling of \ce{H2} fluorescence have shown
that Ly~$\alpha$ emission can dominate the UV spectrum
\citep{Bergin2003,Herczeg2004,Schindhelm2012}.  In this case, the resonant
scattering by atomic hydrogen in the photodissociated regions is also
important, and we treated this similar to \citet{Bethell2011b}.  To increase
the signal-to-noise ratio of line features like Ly~$\alpha$, a smaller
energy is used for photon packets when its frequency falls into the line
profile.  We also include the absorption of UV photons by water.  Since the
abundance of atomic hydrogen and water is affected by chemistry, the whole
process has to be iterated, which is slow but affordable.  A byproduct of the
radiative transfer is the distribution of radiation field over the whole disk,
which will be used as input for chemistry and gas thermal balance.

After establishing the dust temperature distribution, we evolve the disk
chemistry for 1~Myr.  Since the heating and cooling processes are coupled with
chemistry, the gas temperature is evolved in tandem with chemistry based on the
heating and cooling rates.  Namely, we solve the following set of ordinary
differential equations (ODEs)
\begin{equation}
\begin{split}
  \frac{\mathrm{d}}{\mathrm{d}t} X_i &= P_i(X;T) - D_i(X;T), \; i=1,\ldots N, \\
  \frac{\mathrm{d}}{\mathrm{d}t} T &= (\Gamma - \Lambda) / C_\text{v},
\end{split}
\end{equation}
where $X_i$ is the abundance of species $i$, $P_i$ and $D_i$ are the
production and destruction rates of this species, which are functions of the
chemical abundances and temperature (and other physical parameters), and $N$ is
the total number of species.  $C_\text{v}=3k_\text{B}/2$ is the volume-specific
heat capacity of an ideal gas, where $k_\text{B}$ is the Boltzmann constant.  The
exact value of $C_\text{v}$ is not important, because we are only concerned
with the equilibrium temperature, rather than the rate of temperature change.
The heating and cooling rates are contained in $\Gamma$ and $\Lambda$.  We do
not need a separate set of equations to account for the elemental conservation,
since elements are automatically conserved within numerical tolerance.  For
solving the above set of ODEs, we use the DLSODES solver of the
ODEPACK\footnote{\url{http://www.netlib.org/odepack/}} package
\citep{Hindmarsh1983}, which makes use of the sparse structure of the chemical
network.

The initial chemical composition is listed in \reftab{tabChemIni}.  The gas
temperature is set to the dust temperature at $t=0$, and usually reaches steady
state within a short period.  We note that the chemistry cannot always reach a
steady state within 1~Myr, and may still evolve at time scale $\gtrsim10^8$~yr.


We could also solve the chemical equilibrium (or rather quasi-equilibrium) and
thermal equilibrium independently.  But in such an approach iteration for each
single grid point will be needed to achieve a joint convergence, which may pose
some numerical issues and takes more CPU time, while in our approach thermal
equilibrium is guaranteed as far as the heating/cooling time scale is shorter
that the time scale of interest ($\sim$1~Myr).

One note about the global iteration in our code.  For the radiative transfer in
the first iteration, only dust is assumed to be present.  This gives a
distribution of dust temperature and radiation intensity over the whole disk.
Based on this the chemistry and gas temperature is solved on the grid points in
a downward (i.e. from surface to midplane) then outward (from close to the
central star to the disk outer edge) order.  This order has the advantage that
the self-shielding of \ce{H2} and CO can be updated each time a grid point has
been calculated during one iteration.  The radiative transfer is redone before
each chemical and thermal calculation of the whole disk, to take into account
the effects (H scattering, water absorption) due to updated chemical
composition.  The changes in the radial water abundance profile at different
vertical height as the iteration proceeds will be described later in
Section \ref{secHowImpWater}.  Since the code has a Monte Carlo component (for
the radiative transfer) built in, perfect convergence is not expected.

\begin{table}
\centering
\caption{Initial chemical composition, relative to the total number density of
hydrogen nuclei.  $a(b)\equiv a\times10^b$.  \label{tabChemIni}}
\begin{tabular}{ll}
\hline\hline
Species    &  Abundance \\
\hline
\ce{H2 }   &  $0.5$  \\
\ce{He }   &  $0.09$     \\
\ce{CO }   &  $1.4(-4)$  \\
\ce{N  }   &  $7.5(-5)$  \\
\ce{H2O} (ice)  &  $1.8(-4)$  \\
\ce{S  }   &  $8(-8)$  \\
\ce{Si+}   &  $8(-9)$  \\
\ce{Na+}   &  $2(-8)$  \\
\ce{Mg+}   &  $7(-9)$  \\
\ce{Fe+}   &  $3(-9)$  \\
\ce{P  }   &  $3(-9)$  \\
\ce{F  }   &  $2(-8)$  \\
\ce{Cl }   &  $4(-9)$  \\
\hline
\end{tabular}
\end{table}

\subsection{Chemical Network}

We use the full UMIST RATE06 network \citep{Woodall2007} for our gas phase
chemistry.  Details for the implementation of this network can be seen in that
paper.  In addition, we include dissociation of \ce{H2O} and \ce{OH} by
Ly~$\alpha$ photons, adsorption of major gas phase species onto the dust
grain, and desorption of species on the dust grain surface either thermally, or
induced by cosmic-rays and UV photons.  Two-body reactions on the dust grain
surface are also included, leading to the formation of \ce{H2O}, \ce{CH3OH},
\ce{CH4}, etc.  The surface network is taken from \citet{Hasegawa1992}.
Recombination of ions with charged dust grains is included.  In total the
chemical network has 467 species and 4801 reactions.  We describe some
reaction types of special importance in the following.

\subsubsection{Adsorption}

The adsorption rate of species X is
\begin{equation}
  R_\text{ad}(\text{X}) = s \sigma v_\text{T} n_\text{dust},
  \label{eqRad}
\end{equation}
where $s$ is the sticking coefficient, $\sigma$ is the cross section of dust
particles, $v_\text{T}$ is the thermal speed of species X, and $n_\text{dust}$
is the density of dust particles.  We have
\begin{equation}
  \sigma = \pi a^2,\ v_\text{T} = \sqrt{\frac{8k_\text{B} T_\text{gas}} {\pi
  m_\text{X}}},
  \label{eqThermalVel}
\end{equation}
where $a$ is the average radius of dust particles, $k_\text{B}$ is the
Boltzmann constant, and $m_\text{X}$ is the mass of a particle of X.

We calculate the sticking coefficient using a formula from
\citet{Chaabouni2012}
\begin{equation}
  s = \frac{1 +
  T_\text{gas}/(15\tilde{m}_\text{X})}{\left[1+T_\text{gas}/(38.5\tilde{m}_\text{X})\right]^{2.5}},
\end{equation}
where $\tilde{m}_\text{X}$ is the mass number of X.
The numbers in the above formula are interpolated from the parameters for
nonporous amorphous solid water ice and silicate dust in \citet{Chaabouni2012}.
The effect of using this formula rather than the commonly used constant value
of one is most important for atomic H at high temperatures.  If a constant value
is used, the formation of \ce{H2} may heat the gas to unrealistically high
temperatures in the photo\-dissociated region.

\subsubsection{Thermal Desorption}

The thermal desorption rate is
\begin{equation}
  k_\text{evap,th} = \nu e^{-E_\text{des}/T_\text{dust}},
\end{equation}
where $\nu$ is the characteristic vibrational frequency of species X on the
dust grain surface \citep{Hasegawa1992},
\begin{equation*}
  \nu = \sqrt{\frac{2 n_\text{S}
  E_\text{des}} {\pi^2 m_\text{X}}},
\end{equation*} 
$n_\text{S}$ being the number density of surface sites, usually taken to be
$10^{15}$~cm$^{-2}$, and $E_\text{des}$ is the desorption energy of species X,
for which we adopt the values from \citet{Garrod2008}.  Typically $\nu$ is of
the order of $10^{12}$~Hz.

\subsubsection{Photodesorption}

The yield of a species on the dust grain surface per incident UV photon can be
empirically written as \citep{Oberg2009a, Oberg2009b}
\begin{equation}
  Y = (a + b\times T_\text{dust}) (1-e^{-x/l}),
\end{equation}
where $x$ is the thickness of the ice, and $a$, $b$, and $l$ may be
approximated as constants, which are determined experimentally.
The UV flux is calculated in the radiative transfer part of the code.

For \ce{H2O}, CO, and \ce{CO2}, we use the measured value of the $(a,b,l)$
parameters from \citet{Oberg2009a, Oberg2009b}:
\begin{align*}
  \ce{H2O}:\ & 1.3\times10^{-3}, \; 3.2\times10^{-5}, \; 2,\\
  \ce{CO}: \ & 2.7\times10^{-3}, \; 0,                \; 1,\\
  \ce{CO2}:\ & 2\times10^{-3},   \; 0,                \; 3.
\end{align*}
For all the other species, we assume $a=10^{-4}$, $b=0$, and $l=1$.

\subsubsection{Cosmic-ray Desorption}

Cosmic-ray desorption is important for the deep and cold region of the disk.
We adopt the treatment of \citet{Hasegawa1993}, in which a dust grain is
assumed to be episodically heated by cosmic-rays to a high temperature (70~K)
followed by evaporation of a large fraction of its ice mantle.  Namely
\begin{equation} k_\text{evap,CR} = k_\text{evap,th}(T{=}70\;\text{K})\;
f(70\;\text{K}), \end{equation} where $f(70\;\text{K})$ is the fraction of time
for the dust to spend at temperature $\sim$70~K, which is estimated to be
$3.16\times10^{-19}$ for a dust grain size of 0.1~$\mu$m and a total cosmic-ray
ionization rate of ${\sim}10^{-17}$~s$^{-1}$ \citep{Leger1985}.  The cosmic-ray
intensity is attenuated with an $e$-fold column density of 96~g~cm$^{-2}$
\citep{Umebayashi1981}, and the induced evaporation rate is scaled down
accordingly.  A different grain size distribution and cosmic-ray spectrum would
produce different values for these parameters, though many details are subject
to large uncertainties \citep{Cleeves2013}.

\subsubsection{\ce{H2} Formation}

Since H participates in many surface reactions other than the formation of
\ce{H2}, we cannot simply assume that all the H atoms adsorbed onto the dust grain are
converted into \ce{H2} molecule.  So we treat the formation of \ce{H2} on dust
grain surface as a normal two-body reaction between two H atoms.  Hence the
formation rate (i.e. the number of \ce{H2} molecules formed per unit volume per unit time) is
\begin{equation}
  R(\ce{H2}) = k_\text{HH} n^2(\text{H}_\text{gr}),
  \label{eqH2Form}
\end{equation}
where $n(\text{H}_\text{gr})$ is the number density of H atoms on the grain surface, which recombine with a rate coefficient
\begin{equation}
  k_\text{HH} = \frac{\nu_\text{H}} {n_\text{dust}N_\text{S}}
  e^{-E_\text{diff}/T_\text{dust}},
\end{equation}
in which $E_\text{diff}$ is the energy barrier for migrating over the dust
grain surface, usually taken to be half of the desorption energy, $N_\text{S}$ is
the number of sites per dust grain, and $n_\text{dust}$ is the number density
of dust particles.  At low temperatures, quantum tunneling becomes important,
and the exponential part will be replaced by \citep{Hasegawa1992}
\begin{equation*}
  e^{-{2a_\text{diff}}\sqrt{2m_\text{X}E_\text{diff}}/{\hbar}},
\end{equation*}
where $a_\text{diff}$ of the order of 1~\AA{} is the barrier width for surface
migration.  We assume H atoms on dust grain surface are chemisorbed, and set
the desorption energy to $10^4$~K according to \citet{Cazaux2004}.
Physisorption alone is not enough to account for the abundance of \ce{H2} in
the hot regions of the interstellar medium  \citep[ISM;][]{Cazaux2006}.
Note that if we assume all the adsorbed H atoms are converted into \ce{H2},
then $R(\ce{H2})$ simply becomes half of the adsorption rate of H as calculated
from \refeq{eqRad}.

\subsubsection{Photodissociation of \ce{H2O} and OH}

The photodissociation of \ce{H2O} and \ce{OH} by generic ISM UV field are
included in the UMIST RATE06 network.  In addition, we include the dissociation
by Ly~$\alpha$ photons using the cross sections from
\citet{vanDishoeck2006}, with $\sigma_\ce{H2O}=1.2\times10^{-17}$~cm$^{-2}$,
and $\sigma_\ce{OH}=1.8\times10^{-18}$~cm$^{-2}$.  The local UV (including
Ly~$\alpha$) flux in the disk is determined from the radiative transfer.
The shielding effect of \ce{H2O} is included in the radiative transfer.

\subsubsection{Photodissociation of \ce{CO} and \ce{H2}}

The photodissociation of CO is included in the UMIST RATE06 network.  For
\ce{H2}, we use a rate coefficient of $4\times10^{-11}$.  The self-shielding of
CO and \ce{H2} are considered based on the formulation of \citet{Visser2009}
and \citet{Draine1996}, respectively.

\subsubsection{Photodissociation of Other Species}

For other species in the UMIST network, their photodissociation rates are
calculated based on the formula given in \citet{Woodall2007}
\begin{equation}
  G_0 \alpha \exp(-\gamma A_\text{V}),
\end{equation} 
where $G_0$ is the unattenuated UV continuum intensity at each location of the
disk relative to the standard ISM value.  Namely, $G_0$ is calculated from the
stellar spectrum assuming only inverse-square-law dilution.  The attenuation
due to dust and possibly water is included in the $A_\text{V}$ parameter, which
is calculated by comparing the local actual UV field (obtained from Monte Carlo
radiative transfer) with the unattenuated one
\begin{equation}
  A_\text{V} = \max(0,\;-1.086 \ln\left[F_\text{UV,att}/F_\text{UV,unatt}\right]).
\end{equation} 
Such a treatment is similar to \citet{Fogel2011}.

\subsection{Heating and Cooling Processes}

\subsubsection{Photoelectric Heating}

For the heating rate due to small grains and polycyclic aromatic hydrocarbons
(PAHs) we use the formula from \citet{Bakes1994}
\begin{equation}
  \Gamma_\text{pe} = 10^{-24} \epsilon G_0 n_\text{H},
\end{equation}
with unit erg~s$^{-1}$~cm$^{-3}$.
$\epsilon$ is given by
\begin{equation}
\begin{split}
  \epsilon &= \frac{4.87\times10^{-2}}
             {1 + 4\times10^{-3}\left(G_0T^{1/2}/n_\text{e}\right)^{0.73}} \\
           &\quad + \frac{3.65\times10^{-2} (T/10^4)^{0.7}}
                         {1 + 2\times10^{-4}\left(G_0 T^{1/2}/n_\text{e}\right)}.
\end{split}
\end{equation}
The PAH abundance used by \citet{Bakes1994} is $1.6{\times}10^{-7}$
relative to H.  We take into account the effect of dust settling and growth on
this heating rate by scaling down the above rate with a factor equal to the
dust-to-gas mass ratio relative to the ISM value (0.01), though the actual
amount of PAH in disks is uncertain.

\subsubsection{Chemical Heating and Cooling}

Chemical reactions can release or absorb energy.  The heating/cooling due to
chemical reactions can be important for keeping the model self-consistent, and
has been considered in some of the previous works \citep[see,
e.g.,][]{Glassgold1973,Dalgarno1974,Hollenbach1979,Glassgold2012}.  We include
the contribution to energy balance from reactions involving the major abundant
species.  The exothermicity or endothermicity of these reactions are calculated
based on the enthalpy of the formation of the reactants and products using the
following formula
\begin{equation}
  \Delta H = \sum \nu_i\Delta_\text{f}H^\text{o}(i),
  \label{eqReactionEnthalpy}
\end{equation}
where the sum is over the reactants and products in a reaction, $\nu_i$ is the
stoichiometric coefficient (negative for reactants and positive for products),
and $\Delta_\text{f}H^\text{o}(i)$ is the enthalpy of formation of a species.
$\Delta H>0$ means the reaction is endothermic.
The contribution of a reaction to the heating/cooling rate is $k \Delta H$,
where $k$ is the reaction rate.  The enthalpy of formation of chemical species
are slowly changing functions of temperature and pressure.  For our purpose it
suffices to use the values measured at standard condition (i.e. $p=1$~bar,
$T=298$~K).  The thermochemical data are taken from the NIST
webbook\footnote{\url{http://webbook.nist.gov/chemistry/}},
\citet{Binnewies2002}, \citet{Vandooren1991}, and \citet{Nagy2010}.
In total 591 reactions are included to contribute to the gas heating/cooling.

\subsubsection{Heating by H$_2$ Formation}

Similar to \citet{Sternberg1989} and \citet{Rollig2006}, we assume one third of
the energy released ($=4.5$~eV) in the formation of a H$_2$ molecule from
combination of two H atoms are converted into heat of the gas.  The
corresponding heating rate is
\begin{equation}
  \Gamma_{\text{H}_2\;\text{form}} = 2.4\times10^{-12}~\text{erg}~R(\text{H}_2),
\end{equation}
where $R(\text{H}_2)$ (see \refeq{eqH2Form}) is the formation rate of H$_2$ in
unit of cm$^{-3}$~s$^{-1}$.

\subsubsection{Heating by Viscous Dissipation}

We use the usual $\alpha$-prescription.  The heating rate is
\begin{equation}
  \Gamma_\text{vis} = \frac{9}{4} \alpha \rho c_\text{S}^2 \omega_\text{K},
  \label{eqVisHeat}
\end{equation}
where $\rho$ is the mass density of the gas, $c_\text{S}$ is the sound speed,
and $\omega_\text{K}$ is the Keplerian angular velocity.

Usually $\alpha$ is assumed to be a constant of the order of 0.01 -- 1.  As
noted by \citet{Woitke2009}, the heating rate calculated from \refeq{eqVisHeat}
can become unphysical and gives very high temperature (${>}10^4$~K) when the
density is very low.  Hence the calculated high temperature in the top layers
of the disk may not be trusted, though this does not affect our goal of study,
which is focused on the deeper shielded region.  Thus we use the analytical
formula of \citet{Bai2011} fitted from non-ideal magnetohydrodynamical
simulations
\begin{equation}
\begin{split}
  &\alpha = \frac{1/2}{[\left(2500/Am^{2.4} + (8/Am^{0.3} +
  1)^2\right]^{1/2}}, \\
  &Am = n_\text{ion} \beta_\text{ion} / \Omega,
\end{split}
\end{equation}
where $n_\text{ion}$ is the ion density, $\beta_\text{ion}$ is the ion-neutral
collision rate, and $\Omega$ is the local Kepler angular velocity.
At the surface of the disk where the density is low, the ambipolar diffusion
parameter $Am$ will be small and so will $\alpha$ (${\sim}10^{-4}$), which will
partially alleviate the problem of temperature getting unphysically high.

\subsubsection{Heating by Cosmic-ray and X-ray}

The cosmic-ray heating rate is \citep{Bruderer2009}
\begin{equation}
  \Gamma_{\text{CR}} = 1.5\times10^{-11} \times \zeta_\text{CR} n_\text{gas},
\end{equation}
where $\zeta_\text{CR}$ is the cosmic-ray flux, and $n_\text{gas}$ is the gas density.

For the X-ray heating, we calculate the X-ray photoelectric cross sections for
the gas and dust using the interpolation table in \citet{Bethell2011a} assuming
a representative X-ray photon energy of 1~keV (corresponding to a $10^7$~K
black body), and assume that each ion pair release 18~eV into the gas
\citep{Glassgold2012}.  The total X-ray flux from the central star is taken to
be $10^{-3}$~$L_\odot$.  The X-ray intensity at each location is attenuated by
the column towards the central star, similar to \citet{Glassgold2004} (see also
\citealt{Glassgold1997}).  The contribution of X-ray to the ionization rates
are treated similar to \citet{Bruderer2009a}, namely, we enhance the cosmic-ray
ionization rates by the calculated X-ray ionization rates.  As remarked by
\citet{Bethell2011a}, the scattered diffusion of X-ray with energy
${\sim}1-2$~keV is not important, though a scattered hard X-ray field can be
important even deep in the midplane, especially when the cosmic-ray intensity
has been suppressed by the stellar wind of the central star
\citep{Cleeves2013}.  Hence our simple treatment tends to underestimate the
X-ray ionization rate.  \citet{Aresu2011} find that a high X-ray luminosity of
the central star can reduce the amount of hot water vapor by a factor of 20
relative to the case with zero X-ray luminosity.  In our test runs we did not
see such an effect, and we interpret this as (see also \citealt{Meijerink2012})
due to the fact that our parameterized model does not adjust the vertical
structure accordingly when the heating rate is increased, which would otherwise
decrease the opacity to the dissociating UV photons.

\subsubsection{Energy Exchange by Gas-dust Collision}

The energy exchange between gas and dust particles due to collisions can heat
or cool the gas, depending on whether the gas temperature is lower or higher
than the dust.  The energy exchange per unit volume per unit time is
\citep{Hollenbach1979}
\begin{equation}
  \Lambda_\text{gd} = 2k_\text{B} (T_\text{gas} - T_\text{dust}) v_\text{T}
  \sigma_\text{d} n_\text{dust} n_\text{gas} f_\text{a},
  \label{eqCoolingGD}
\end{equation}
where the factor two is due to the fact that more energetic particles collide
with the dust grain more frequently, $v_\text{T}$ is the thermal speed of gas
particles (\refeq{eqThermalVel}), $\sigma_\text{d}$ is
the mean cross section of dust particles, and $f_\text{a}$ is the accommodation
coefficient, for which we take the expression from \citet{Hollenbach1989}
\begin{equation}
  f_\text{a} = 1 - 0.8 e^{-75/T_\text{gas}}.
\end{equation}
If more than one dust species exists, their contributions are added together.
We take into account the possibility that one type of dust particle heats the
gas while another cools the gas.

\subsubsection{Other Heating and Cooling Mechanisms}

\paragraph{Heating by photodissociation of \ce{H2}}
We use the formula from \citet{Tielens2005}
\begin{equation}
  \Lambda_{\text{ph},\ce{H2}} = 1.36\times10^{-23} G_0 n(\ce{H2}),
\end{equation}
where the unit is erg~s$^{-1}$~cm$^{-3}$.
$G_0$ is the local UV intensity with self-shielding and dust extinction
taken into account.

\paragraph{Heating by photodissociation of \ce{H2O} and \ce{OH}}
Their contributions to heating are calculated with
\begin{equation}
\begin{split}
  \Gamma_{\text{ph},\ce{H2O}} &= F_{\text{Ly}\alpha} \sigma_{\ce{H2O}}
  n(\ce{H2O}) E_{\ce{H2O}}, \\
  \Gamma_{\text{ph},\ce{OH}} &= F_{\text{Ly}\alpha} \sigma_{\ce{OH}}
  n(\ce{OH}) E_{\ce{OH}},
\end{split}
\end{equation}
where $F_{\text{Ly}\alpha}$ is the local Ly~$\alpha$ number flux, $\sigma$
is the photodissociation cross section, and we take
$E_\ce{H2O}=8.1\times10^{-12}$~erg and $E_\ce{OH}=9.2\times10^{-12}$~erg,
obtained by subtracting from the Ly~$\alpha$ photon energy the enthalpy
change of the two dissociation reactions.  $E_\text{H2O}$ may overestimate the
actual value by a factor of 2 -- 4, since a portion of the absorbed energy can
be used to excite the internal modes of OH \citep{Mordaunt1994}.

\paragraph{Heating by ionization of atomic carbon}
We use the formula from \citet{Tielens2005}
\begin{equation}
  \Gamma_\text{ion,C} = 2.2\times10^{-22} G_0 n(\ce{C}),
\end{equation}
with unit erg~s$^{-1}$~cm$^{-3}$.

\paragraph{Cooling by electrons recombine with small dust grains}
We use the analytical formula from \citep{Bakes1994}
\begin{equation}
\begin{split}
  \Lambda_\text{recom} &= 3.49\times10^{-30}\ n_\text{e} n_\text{H}\ T^{0.944}\\
  &\quad \times \left(\frac{G_0 T^{1/2}}
  {n_\text{e}}\right)^{\beta},
\end{split}
\end{equation}
with unit erg~s$^{-1}$~cm$^{-3}$.
$\beta = 0.735/T^{0.068}$.  This cooling rate is reduced when dust is
depleted.

\paragraph{Cooling by the rotational transitions of \ce{H2}, and the rotational
and vibrational transitions of CO and \ce{H2O}} We calculate these cooling
rates using the interpolation tables from \citet{Neufeld1993} and
\citet{Neufeld1995} rather than direct radiative transfer for the sake of
speed.  These tables were made based on the escape probability approximation,
in which a velocity gradient is needed, for which we use the radial gradient of
the orbital speed
\begin{equation*}
{\mathrm{d}v}/{\mathrm{d}r} =
\frac{1}{2}\left(\frac{G M}{r^3}\right)^{1/2},
\end{equation*} 
though note that since the
orbital velocity is perpendicular to the radial direction, the photons cannot
easily escape in the radial direction, but rather at an angle with it.

\paragraph{Heating and cooling by the vibrational transitions of \ce{H2}, and
cooling by \ce{C+} and O emission}
We use the analytical formulae from \citet{Rollig2006}.  The escape probability
for the \ce{C+} and O lines are also calculated based on the radial gradient of the
orbital speed and the local velocity dispersion.

\paragraph{Cooling by Ly~$\alpha$ emission, free-bound, and free-free
emissions}
They are usually not important for our purpose.  We include them for
completeness, using the formulae in \citet{Tielens2005} and \citet{Draine2011}.

\subsection{Disk Structure}

We assume that the disk is static in the vertical direction and is in Keplerian
rotation in the azimuthal direction (only needed for the line radiative transfer).
The axisymmetric disk density structure we use takes the following
parameterized form in cylindrical coordinates $(r,z)$ \citep{LyndenBell1974,
Hartmann1998, Andrews2009, Cleeves2013}
\begin{equation}
  \rho(r, z) = \frac{\Sigma}{\sqrt{2\pi}h}
    \exp\left[-\frac{1}{2}\left(\frac{z}{h}\right)^2\right],
\end{equation}
where
\begin{equation}
\begin{split}
  \Sigma(r)  &= \Sigma_\text{c} \left(\frac{r}{r_\text{c}}\right)^{-\gamma}
  \exp\left[-\left(\frac{r}{r_\text{c}}\right)^{2-\gamma}\right], \\
  h          &= h_\text{c} \left(\frac{r}{r_\text{c}}\right)^\psi.
\end{split}
\label{eqSigmar}
\end{equation}
The disk mass (gas or dust) is
\begin{equation}
\begin{split}
  M_\text{disk} &= \int_{r_\text{in}}^{r_\text{out}} \Sigma\; 2\pi r \mathrm{d}r \\
  &=\frac{2}{2-\gamma}
  \pi r_\text{c}^2 \Sigma_\text{c}
  \left[e^{-\left(\frac{r_\text{in}\vphantom{_g}}{r_\text{c}}\right)^{2-\gamma}} -
  e^{-\left(\frac{r_\text{out}\vphantom{_g}}{r_\text{c}}\right)^{2-\gamma}}\right].
\end{split}
\label{eqSigmac}
\end{equation}
A list of the parameters involved and their meanings are in \reftab{tabDiskParams}.
Note that $\Sigma_\text{c}$ is not included as an independent parameter since
it can be calculated from \refeq{eqSigmac}.  Also note that the gas and dust
components of the disk can each have a different set of values for these
parameters.

\begin{table}[htbp]
\centering
\caption{Major parameters in our model, and their fiducial values.
\label{tabDiskParams}}
\begin{tabular}{lp{0.8\linewidth}}
\hline\hline
\multicolumn{2}{c}{Stellar parameters} \\
$M_\text{star}$ & Mass; 0.6~$M_\odot$ \\
$R_\text{star}$ & Radius; 1~$R_\odot$ \\
$T_\text{star}$ & Effective temperature; 4000~K \\
$L_\text{star}$ & Total luminosity; 0.25~$L_\odot$ \\
$L_\text{UV}$ & UV continuum luminosity; 0.02~$L_\odot$ \\
$L_{\text{Ly}\alpha}$ & Ly $\alpha$ luminosity; 0.004~$L_\odot$ \\
$L_{\text{Xray}}$ & X-ray luminosity; 0.001~$L_\odot$ \\
\hline
\multicolumn{2}{c}{Disk parameters} \\
$r_\text{in}$  &  radius of the disk inner edge; 1 AU\\
$r_\text{out}$ &  radius of the disk outer edge; 140 AU\\
$M_\text{disk}$ & disk gas mass; $0.05~M_\odot$\\
$M_\text{dust}$ & disk dust mass; $0.01~M_\text{disk}$\\
$r_\text{c}$   &  a characteristic radius; 100 AU \\
$h_\text{c}$   &  scale height at the characteristic radius; 10 AU\\
$\gamma$    & power index for the disk surface density distribution; 1\\
$\psi$      & power index for the scale height as a function of radius; 1\\
\hline
\multicolumn{2}{c}{Other parameters} \\
$\alpha$ & Turbulent viscosity parameter; 0.01 \\
$\zeta$ & Cosmic-ray ionization rate; \newline $1.36{\times}10^{-17}$~s$^{-1}$
\\
$G_{0,\text{ISM}}$ & ISM UV field intensity; 1 \\
\hline
\end{tabular}
\end{table}

\section{Results}
\label{secResults}

\subsection{A Fiducial Model}
\label{secFiducial}

We first show a fiducial model with stellar and disk parameters listed in
\reftab{tabDiskParams}.  The disk is assumed to have an inner hole with a sharp
edge.  Except for the radius of this edge, which is set to 1~AU here, those
parameters are set to mimic the transition disk TW~Hya as derived by
\citet{Calvet2002}.  The input stellar UV spectrum (including the
Ly~$\alpha$ line emission) is the observed spectrum of TW~Hya
\citep{Herczeg2002,Herczeg2004}.  We did not take into account the possible UV
variability.  For the initial chemical composition, we assume hydrogen is in
\ce{H2}, carbon is in gas phase CO, and oxygen not in CO is in water ice, while
all the other elements are in atomic form.  We assume two types of dust grains,
each with a MRN (Mathis-Rumpl-Nordsieck) size distribution \citep{Mathis1977}.
The two dust components are assumed to be spatially coexistent in this fiducial
model.  The larger population has $r_\text{min}=1$~$\mu$m and
$r_\text{max}=100$~$\mu$m, with a dust-to-gas mass ratio of 0.01, while the
smaller population has $r_\text{min}=0.01$~$\mu$m and $r_\text{max}=1$~$\mu$m,
with a dust-to-gas mass ratio of $2{\times}10^{-5}$.  Larger values for
$r_\text{max}$ of the big grains has been used in the literature for fitting
the disk spectral energy distribution.  However, the chemical processes mainly
depends on the total available dust grain surface area, which is more sensitive
to the assigned overall mass fractions of the small and big grains than the
value of $r_\text{max}$.  The dust material is assumed to be a $7:3$ mixture of
``smoothed UV astronomical silicate'' and graphite.  The optical parameters of
the dust are taken from the website of Bruce~T.~Draine%
\footnote{\url{http://www.astro.princeton.edu/~draine/dust/dust.diel.html}}
\citep{Draine1984,Laor1993}.

\begin{figure}[htbp]
\includegraphics[width=\linewidth]{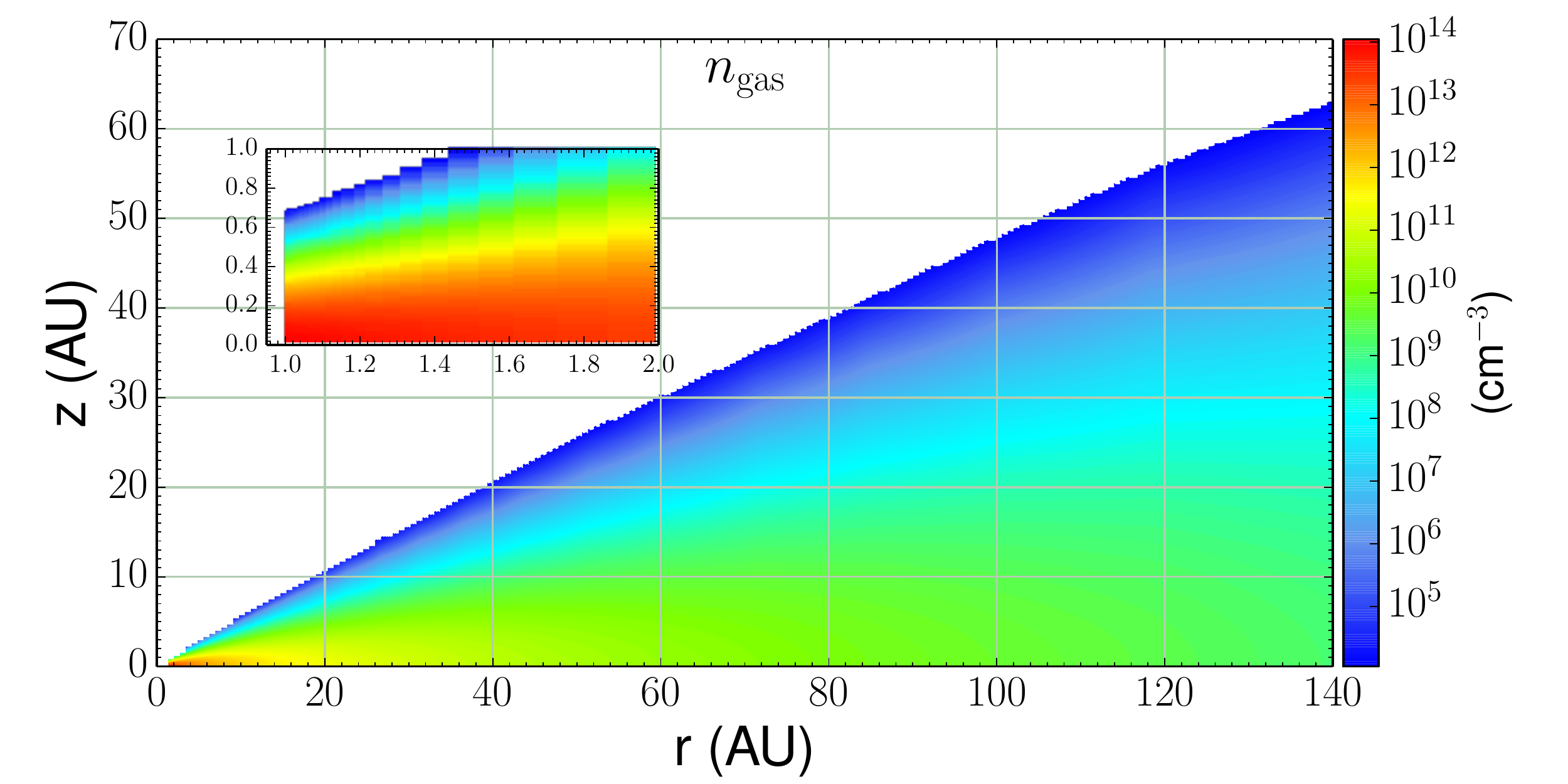}\\
\includegraphics[width=\linewidth]{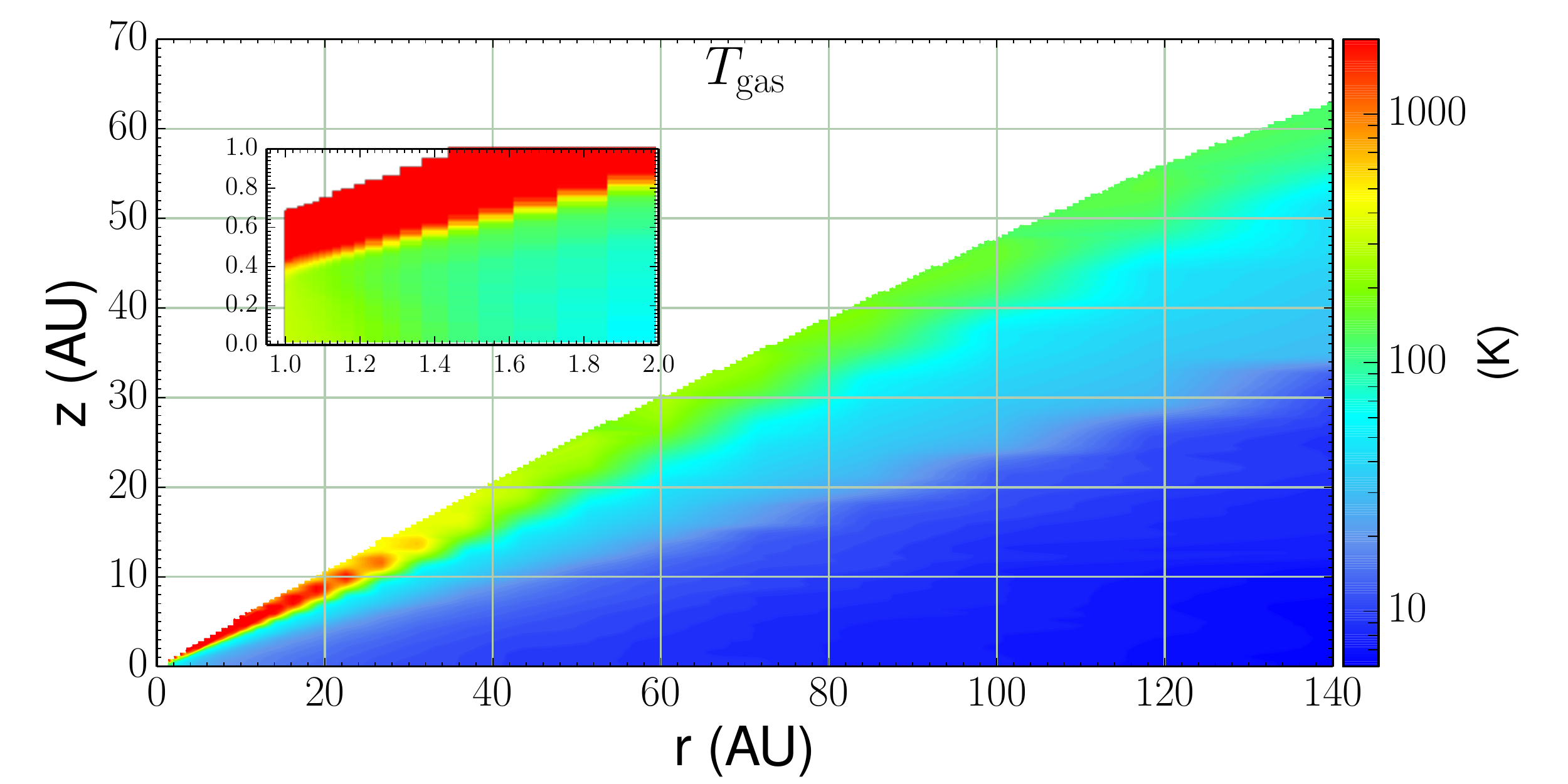}\\
\includegraphics[width=\linewidth]{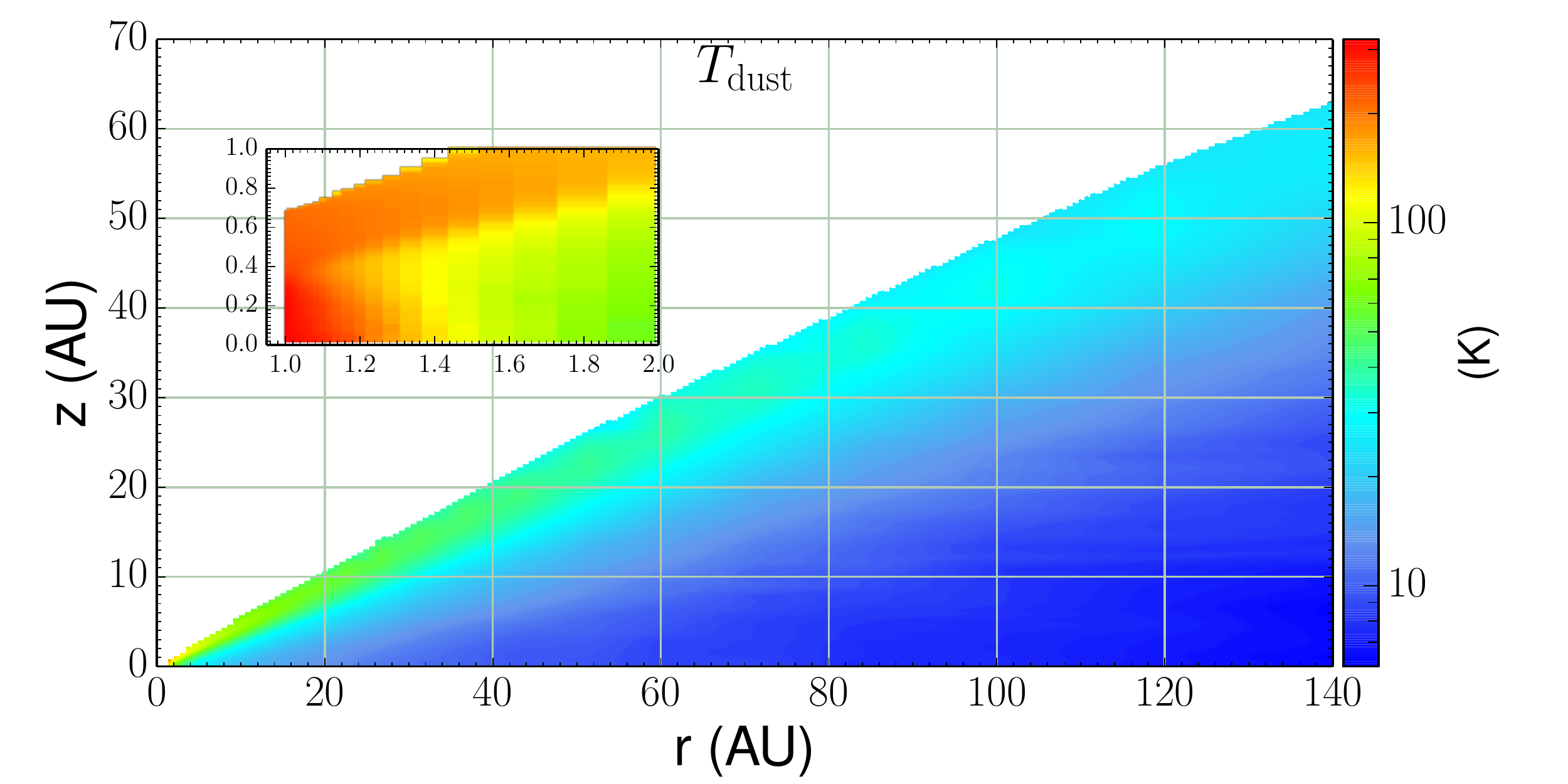}
\caption{Distribution of gas density (top) used as input, and the calculated
distribution of gas (middle) and dust (bottom) temperature in the fiducial
model.
\label{figGasTemperature}}
\end{figure}

\begin{figure}[htbp]
\includegraphics[width=\linewidth]{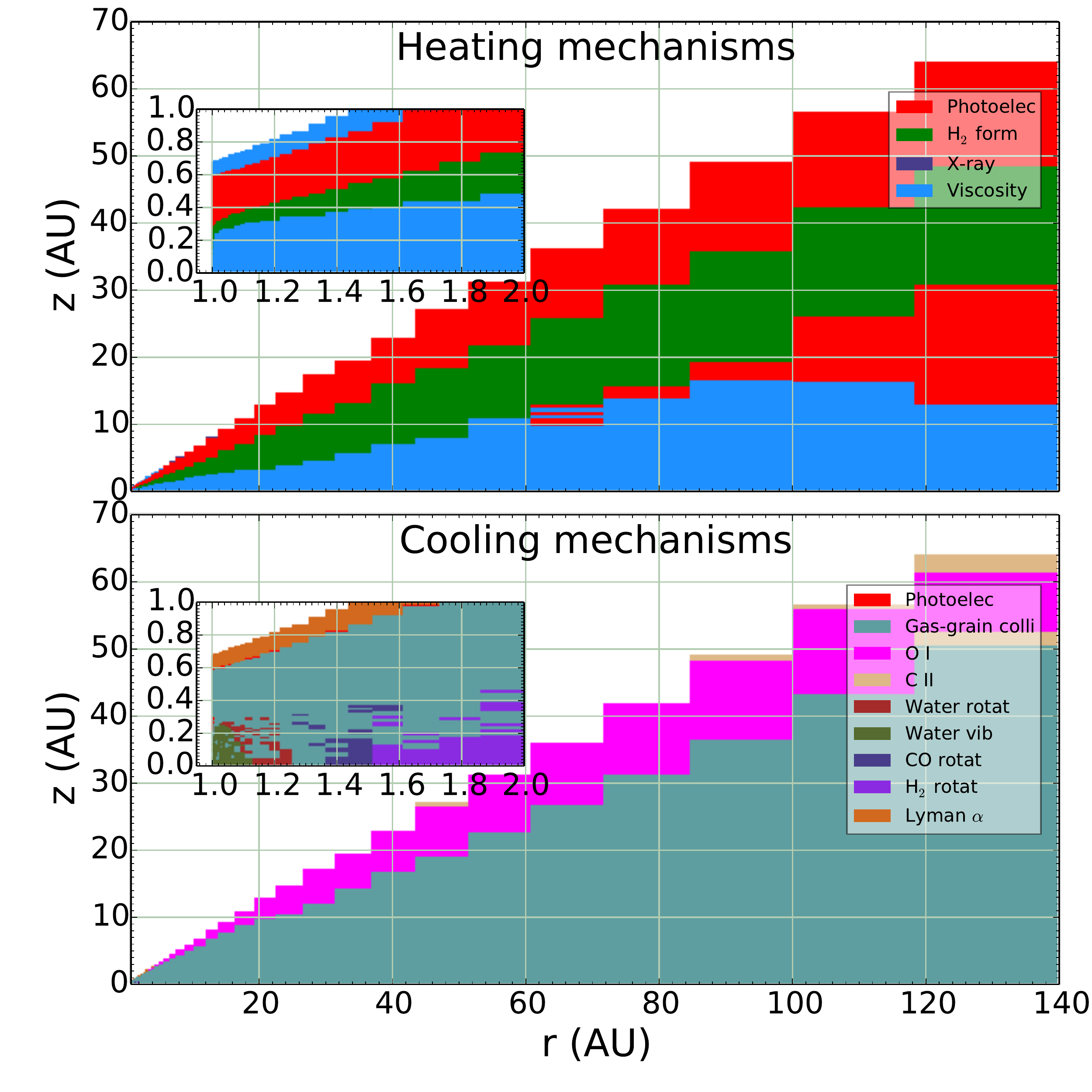}
\caption{Dominant heating and cooling mechanisms in the disk.  Note that at
each location multiple heating/cooling mechanisms can be important, while only
the one contributes most is drawn, which makes the distribution appear
``sporadic''.
\label{figHeatingCooling}}
\end{figure}

The input density structure, and the gas and dust temperature distribution
obtained from radiative transfer and thermal balance calculation are shown in
\reffig{figGasTemperature}.  The dominant heating and cooling mechanisms are
shown in \reffig{figHeatingCooling}.  In the disk upper layer heating is
dominated by photoelectric effect, followed by \ce{H2} formation heating in the
photodissociation layer, and viscous heating in the deep region.  Cooling is
dominated by O~\textsc{i} and C~\textsc{ii} lines in the upper layer, and by
accommodation on the dust grains in the lower dense layers.  Overall the
distribution of the dominant heating and cooling mechanisms is similar to what
is shown in \citet{Woitke2009}.

\begin{figure}[htbp]
\includegraphics[width=\linewidth]{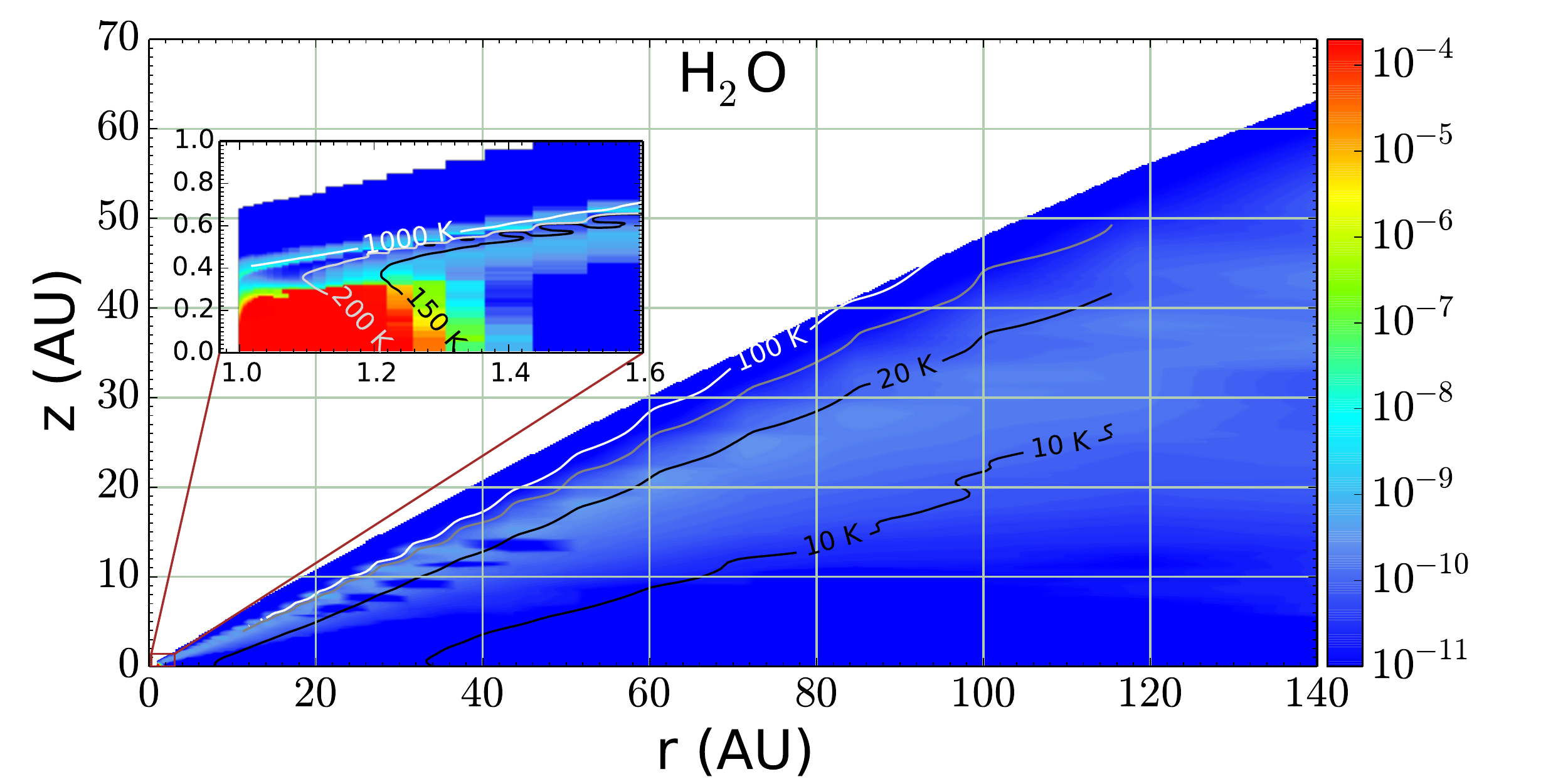}
\caption{Distribution of water vapor in the disk, overlaid with contours of gas
temperature.  The inset show zoom-in view of the distribution close to the
inner edge at 1~AU.
\label{figWaterDistr1AU}}
\end{figure}

The gas phase \ce{H2O} distribution is shown in \reffig{figWaterDistr1AU}.
Hot/warm water with temperature $\gtrsim$200~K is concentrated in a very small
region close to the inner edge near the midplane, which can be seen clearly in
the inset of this figure.  The abundance of water vapor in this region is
$\gtrsim2\times10^{-4}$, which means essentially all the oxygen not in CO are
found in \ce{H2O} gas.  The total warm water mass is $7.7\times10^{26}$~g,
which is equivalent to 560 times the mass of Earth oceans ($M_\text{ocean}$),
or $\sim$0.1~$M_\text{Earth}$.
Although the UV flux from the star is very strong ($G_0\sim10^7$) at the disk
inner edge, the destruction of water by UV radiation is completely quenched
only slightly outward, due to the high density in this region.  For example,
with $n_\text{gas}=10^{14}$~cm$^{-3}$ and a dust-to-gas mass ratio of 0.01
(assuming 1~$\mu$m for the dust grain radius), the attenuation length of the UV
field is of the order of $10^{-4}$ AU, which is too small to be seen in
\reffig{figWaterDistr1AU}.  This explains why the warm water is located so
close to the inner edge, as envisaged by \citet{Cleeves2011}.  Similar
distributions can also be found in \citet{Woitke2009a}, \citet{Aresu2011},
\citet{Heinzeller2011}, and \citet{Meijerink2012}.

To understand the small extent of the warm water distribution, it suffices to
know that, in a well shielded region, a high abundance of gas phase water can
only be maintained if the dust temperature is higher than $\sim$150~K, the
evaporation temperature of water, otherwise water will condensate onto the dust
grains to form ice.  The radial temperature gradient close to the inner edge is
very steep (see \reffig{figTAnafitting}), due to scattering and re-emission of
radiation towards the disk upper and lower surface, which significantly reduces
the amount of energy propagating radially outward.  A semi-analytical account
of the dust temperature profile in the midplane based on the diffusion
approximation of radiative transfer is in Appendix~\ref{secAnaTProfile}, where we will see that the dust temperature drops from $>$300~K at
the edge (1~AU) to $100-150$~K at $r=1.5$~AU, beyond which water can only exist
as ice unless some nonthermal desorption mechanisms come into play.

\begin{figure}[htbp]
\centering
\includegraphics[width=\linewidth]{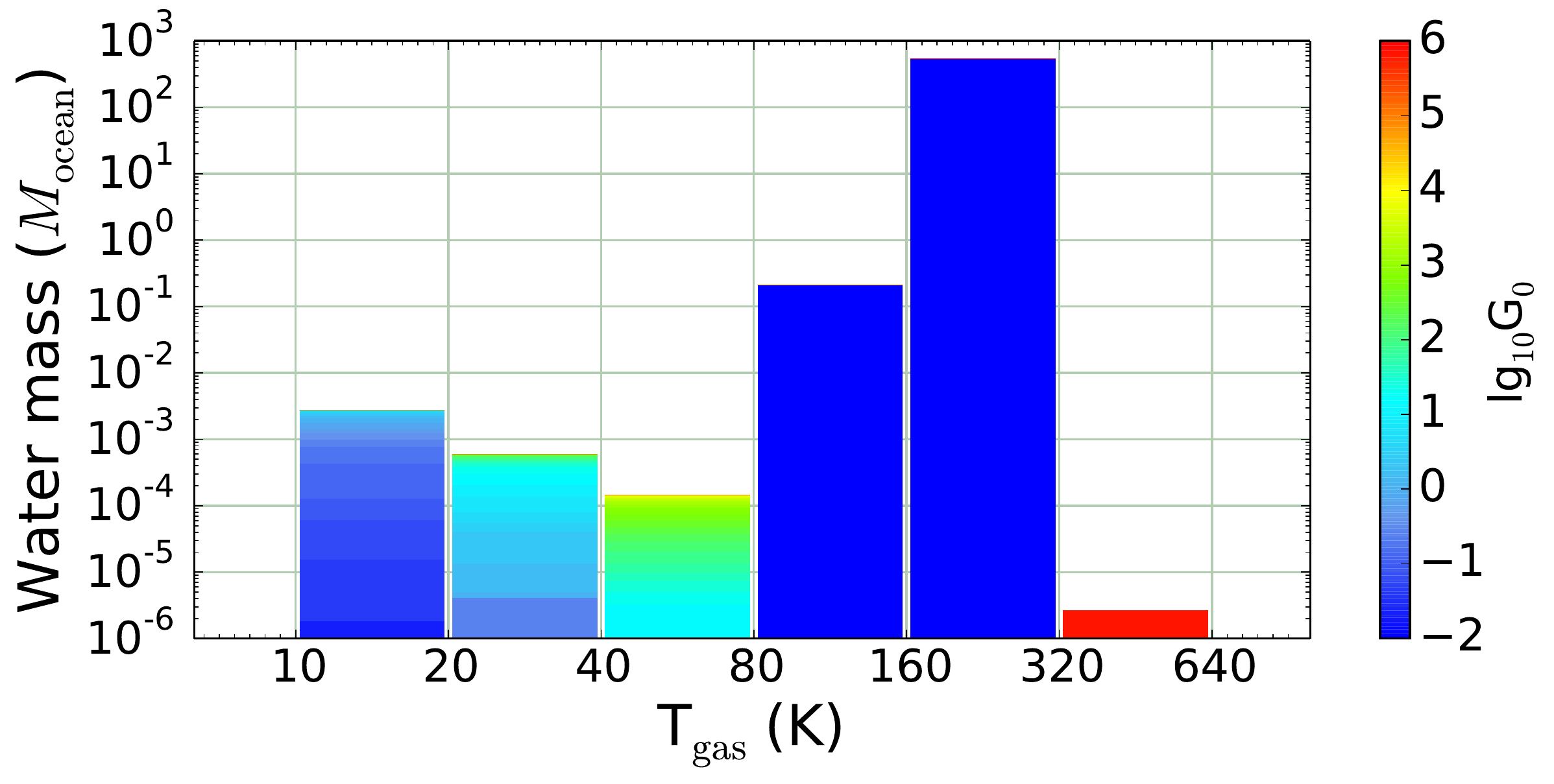}
\includegraphics[width=\linewidth]{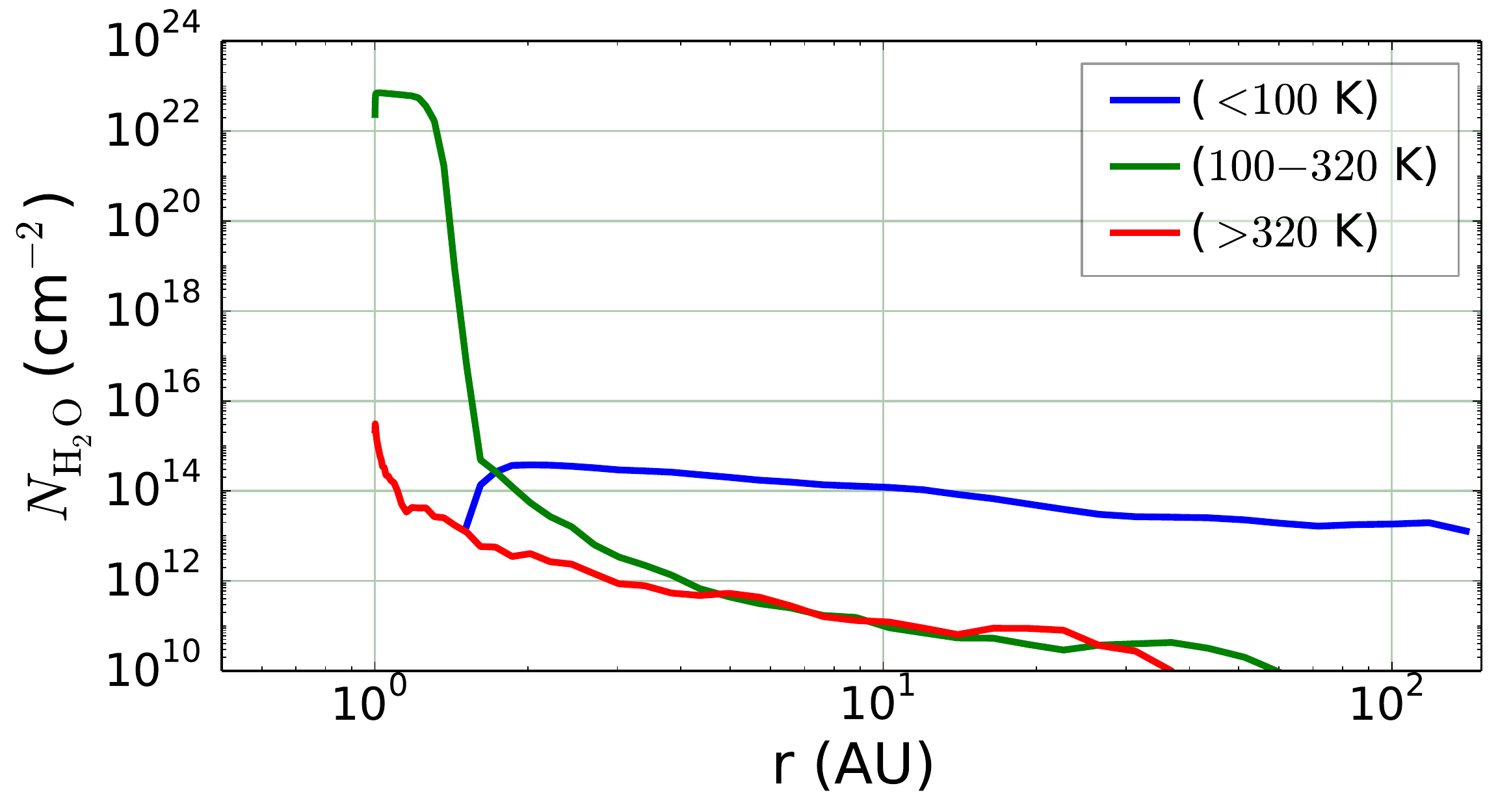}
\caption{Top: water vapor mass in each logarithmic temperature bin in the
fiducial model.  The color scale indicates the UV intensity ($G_0$) of the disk
locations falling into each bin.  Bottom: radial profile of the vertically
integrated column density of water vapor in different temperature ranges.
\label{fighistTgasH2O}}
\end{figure}

Besides the warm water close to the disk inner edge, there is also cold
($\lesssim$50~K) gas phase water in the higher layers throughout the disk (main
panel of \reffig{figWaterDistr1AU}).  The dust temperature at places where this
cold water resides in is well below the evaporation temperature of water, and
here the gas phase water is formed from photodesorption of ice
\citep{Dominik2005}.  The total mass of this diffuse cold water vapor is
$\sim$0.006~$M_\text{ocean}$, but this value depends on the disk size.  The
overall water vapor mass budget as a function of gas temperature can be seen in
the top panel of \reffig{fighistTgasH2O}, from which it is clear that a small
amount of cold ($<$80~K) water is associated with reduced but nonzero UV field
(indicated by the color scale of the histograms).  We note that the total mass
of cold water vapor from our model is close to what was derived by
\citet{Hogerheijde2011}.  A small amount of water vapor can also be produced
through the radiative association reaction
$\ce{H + OH -> H2O + h\nu}$
in the partially
photodissociated layer in the outer disk (see also \citealt{Kamp2013}).
Assuming the abundance of water vapor in the outer cold region is determined
mainly by photodesorption, photodissociation, and adsorption, and assuming the
dust grain is \emph{fully} covered by water ice, we have
\begin{equation}
  X[\ce{H2O}] = \frac{F_\text{UV}\sigma_\text{d} Y \eta_\text{n}} {F_\text{UV} \sigma' +
  n_\ce{H} \eta_\text{n}\sigma_\text{d}v_\text{T}},
\label{eqXH2O}
\end{equation} 
where $F_\text{UV}$ is the UV flux, $\sigma_\text{d}$ is the dust grain cross
section, $Y$ is the photodesorption yield, $\eta_\text{n}$ is the dust-to-gas number
ratio, $\sigma'$ is the water photodissociation cross section, and $v_\text{T}$
is the thermal speed.
As noted by \citet{Dominik2005} and \citet{Bergin2010}, when adsorption is
unimportant, the abundance of water vapor is independent of the UV flux,
\begin{equation}
\begin{split}
  &X[\ce{H2O}] \simeq \frac{\sigma_\text{d} Y \eta_\text{n}} {\sigma'} \\
  = & 2.6{\times}10^{-9}\left(\frac{r_\text{d}}{0.1\;\micron}\right)^2\left(\frac{\eta_\text{n}}{10^{-13}}\cdot\frac{Y}{10^{-3}}\right).
\end{split} 
\label{eqXH2Oapprox}
\end{equation} 
Deeper into the disk, the density becomes higher and the UV flux becomes
weaker, and \refeq{eqXH2Oapprox} will overestimate the water vapor abundance
with respect to \refeq{eqXH2O}.

Also seen in \reffig{fighistTgasH2O} is the existence of a small amount of hot
water vapor ($\gtrsim$300~K), which is similar to the result of
\citet{Woitke2009a}.  Except for the region close to the inner wall, where the
dust temperature is higher than 300~K, the hot water vapor mainly exists in the
upper layer of the disk with $r\lesssim30$~AU, where the density is low enough
that the cooling by accommodation on dust grains is ineffective.  The abundance
of hot water in such region is ${\sim}10^{-10}$, determined by the balance
between photodissociation and warm neutral chemistry (see
\citealt{Woitke2009a}; see also Appendix~\ref{secAnaTre}).

In this section we have seen that the water budget is controlled by a few
mechanisms: dissociation by UV photons, adsorption onto dust grains, and the
shielding by dust (and possibly self-shielding).  Warm water will be preserved
in the gas phase if there is enough shielding while keeping the dust
temperature higher than the water condensation point.  In the next sections we
will discuss their roles in more detail.

\subsection{Effect of Disk Morphology}

\begin{figure}[htbp]
\includegraphics[width=\linewidth]{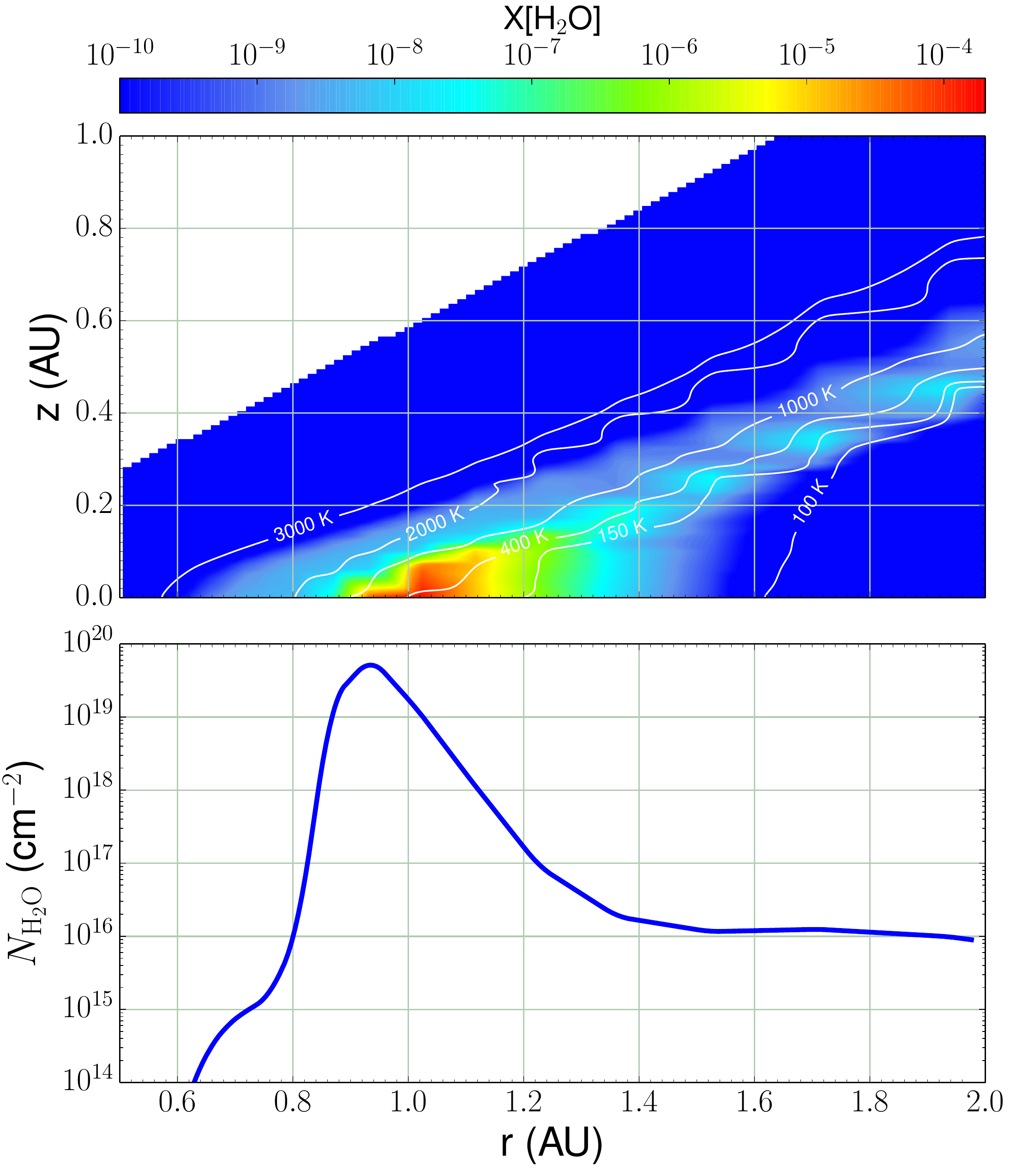}
\caption{Top: distribution of water vapor in the inner disk, overlaid with
contours of gas temperature.  The disk inner edge is at 0.5~AU, with an
exponential taper starting inward from 2~AU.  Bottom: radial profile of water
vapor column density.
\label{figWaterDistr0d5AU_taper}}
\end{figure}

We have assumed a razor sharp inner edge in the fiducial model, which gives a
well confined distribution of warm water close to the inner edge.
We also run a test model with a ``softer'' inner edge, namely, we include an
exponential taper so that the surface density profile becomes
\begin{equation}
  \Sigma'(r) = \Sigma(r) e^{(r-r_0)/r_\text{s}},\ \text{if}~r<r_0,
\end{equation}
where $\Sigma(r)$ is defined in \refeq{eqSigmar}, and $r_0$ and $r_\text{s}$ are the
taper parameters.  In the test model we let $r_\text{in}=0.5$~AU, $r_0=2$~AU,
and $r_\text{s}=0.2$~AU, while other parameters are the same as the fiducial
model.  As shown in the top panel of \reffig{figWaterDistr0d5AU_taper}, warm
water in this test model is confined in a small region around 1~AU, instead of
being close to the edge at 0.5~AU, simply because the dust shielding only
becomes important at $\sim$1~AU due to the reduced density near the inner
edge.  Further out from $\sim$1.5~AU the dust temperature becomes low enough
for water to condense out.  In this test case the total mass of warm water is
very small, only $\sim$0.8~$M_\text{ocean}$.  There are two reasons for this:
the tapered region has much smaller density (hence less mass) than the fiducial
one, and the dust shielding in the vertical direction is also reduced (hence
smaller volume for water to reside in).  The column density of water vapor has
a peak value of a few times $10^{19}$~cm$^{-2}$ at $r\simeq1$~AU, with gas
temperature $\gtrsim$400~K, which are similar to the fitting results of
\citet{Salyk2011}.  For our model to be more close to reality, we will need to
solve the disk physical structure self-consistently in a way similar to, e.g.,
\citet{Nomura2002} or \citet{Woitke2009}, and this will be part of our future
study.

\subsection{Distribution of Warm Water as a Function of the Size of the Disk
Inner Edge}

The size of the inner hole of a protoplanetary disk may evolve as a result of
material exhaustion due to photo-evaporation \citep{Gorti2009,Owen2010}, or accretion
onto the central star or onto the forming planets.  Since the warm water is
concentrated close to the inner wall as seen in the previous section, we expect
the amount of warm water vapor will also evolve as the inner hole expands.

\begin{figure}[htbp]
\includegraphics[width=\linewidth]{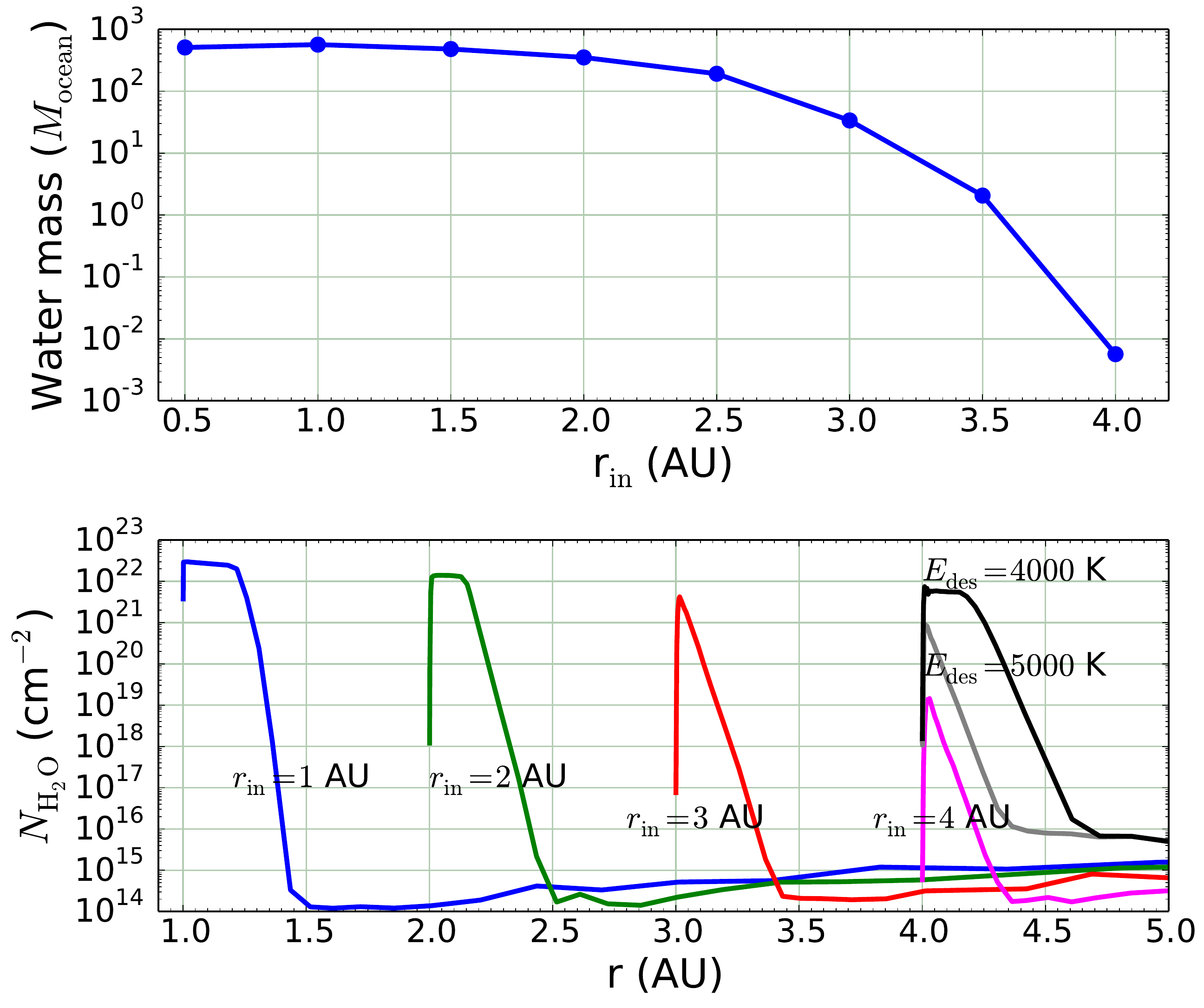}
\caption{
Top: Total mass of warm water (the unit is the mass of Earth ocean
$=1.4{\times}10^{24}$~g) as a function of the size of the disk inner edge.
Bottom: Vertically integrated column density of water vapor for different disk
models with $r_\text{in}$ from 1~AU to 4~AU.  The gray and black curves show test runs
with the desorption energy of water set to 5000~K and 4000~K, respectively,
while all the other curves
are obtained with 5700~K.
\label{figWatervsrin}}
\end{figure}

We run a set of models with all the parameters except $r_\text{in}$ taking the
same value as in the fiducial model.  The mass of water as a function of
$r_\text{in}$ is shown in the top panel of \reffig{figWatervsrin}.  As
expected, the warm water mass generally decreases as $r_\text{in}$ increases.
For $r_\text{in}>3.5$~AU, the amount of warm water vapor becomes very small
because with a central star with bolometric luminosity of $0.25$~$L_\odot,$ the
temperature of the disk wall at $3.5$~AU is ${\sim}170$~K (taking into account
the re-absorption of the radiation emitted by the wall), only marginally higher
than the water condensation temperature, hence warm water vapor can only exist
in a thin skin of the inner wall.

The bottom panel of \reffig{figWatervsrin} shows the vertical column density of
water for $r_\text{in}$ from 1 to 4~AU.  The distribution of water vapor in the
inner disk with column densities $10^{19}$ -- $10^{22}$~cm$^{-2}$ is consistent
with the observations of \citet{Salyk2011}, though the column densities they
derived are mostly in a lower $10^{18}$ -- $10^{19}$~cm$^{-2}$ range.  The
water abundance is ultimately limited from above by the total amount of oxygen
available.  The narrow rings of water vapor with width $\sim$0.2~AU resembles
what was found by \citet{Zhang2013}, though for TW~Hya, they apparently found a
much higher column density (${\sim}10^{22}$~cm$^{-2}$, see their Figure 4)
compared with $10^{19}$~cm$^{-2}$ for the $r_\text{in}=4$~AU case in
\reffig{figWatervsrin}.  One simple way to get a higher column density from our
model is to assume a higher surface density at the inner edge, though in our
present model the surface density at the edge is already high
($\sim$600~g~cm$^{-2}$). One could in principle get a higher oxygen abundance
(hence higher water column) relative to hydrogen if the latter has been
photo-evaporated, without letting the surface density too high.  Another
possibility to increase the water abundance is related to the details of water
adsorption and evaporation (onto and from the dust grains).  Since the dust
temperature at 4~AU is low ($\sim$160~K), water starts to condense out.  The
exact temperature for this to happen depends on the dust properties.  We have
assumed a water desorption energy of 5700~K \citep{Fraser2001}.  Lowering this
value will release more water into the gas phase.  For example, as seen in the
gray curve in the bottom panel of \reffig{figWatervsrin}, reducing it to 5000~K
in a test run gives a peak water vapor column density of $10^{21}$~cm$^{-2}$.
The desorption energy cannot be too low ($\lesssim$4000~K) either, because that
will tend to overproduce gaseous water, as seen in the black curve in
\reffig{figWatervsrin}.
The adsorption and desorption dynamics of dust grains is a complicated issue,
which involves the chemical composition, morphology, as well as crystalline
structure of the ice, which themselves are related to their history of
formation.  The value of 5700~K is appropriate for high density amorphous water
ice.  Mixing with CO and \ce{CO2} ice \citep{Tielens1991,Pontoppidan2008} can
reduce the desorption energy of water due to weaker bonding \citep{Cuppen2007},
though this is not supposed to happen close to the inner edge of the disk,
where CO should be mainly in gas phase.  Finally, we must caution that
potential degeneracy in parametric fitting of the molecular abundance
distribution may render the above comparison premature.

\subsection{The Effect of Stellar Luminosity}

\begin{figure}[htbp]
\includegraphics[width=\linewidth]{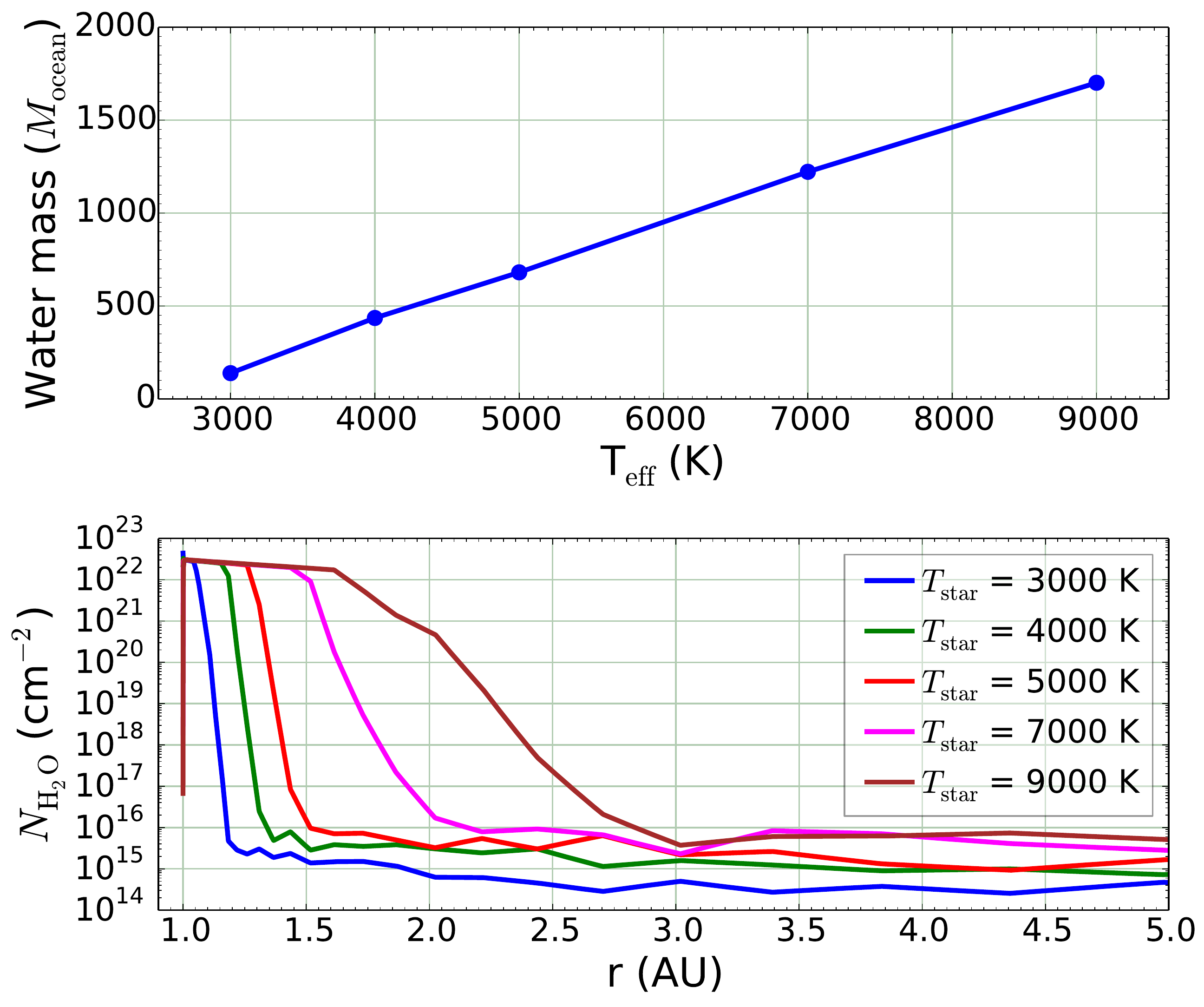}
\caption{
Top: Total mass of warm water as a function of the stellar temperature.  The
stellar spectrum is assumed to be blackbody.  The inner edge is at $r=1$~AU.
Bottom: Vertically integrated column density of water vapor for different
assumed stellar temperatures.
\label{figWatervsTstar}}
\end{figure}

A more luminous central star can warm up the disk, which tends to have more
water molecules released into the gas phase, at the same time it also emits
more UV photons that can dissociate water molecules (though note that for young
stars the UV photons are mainly produced nonthermally by magnetospheric
accretion).  To explore how the stellar luminosity affect the water
distribution, we take a very simple approach by running a set of models with
different effective temperature for the central star, assuming the stellar
emission to be blackbody without any excess, while keeping all the other
parameters the same as in the fiducial model.  The mass of warm water vapor as
a function of stellar temperature can be seen in the top panel of
\reffig{figWatervsTstar}, which shows that the warm water mass increases almost
linearly with stellar temperature.  The reason is that in our model a higher
stellar luminosity increases the overall temperature of the entire disk.  The
higher dissociating photon flux associated with a higher $T_\text{eff}$ does
not necessarily reduce the water content, because UV photons are readily shielded by the
dust, and get converted into photons with lower frequency and no dissociating
capability and diffuse into the disk to warm up the dust and gas.  The column
density profile for different stellar temperatures can be seen in the bottom
panel of \reffig{figWatervsTstar}.  Higher stellar temperature leads to a wider
profile, while the peak column densities are identical, since it is limited by
the surface density.  Admittedly, our treatment here is simplistic.  In
reality, a higher overall temperature would ``blow up'' the disk, reducing the
opacity and letting the UV photons penetrate deeper into the disk to dissociate
the water molecules, at the same time more water molecules might be liberated
into the gas phase due to higher flux of desorbing UV photons.

\begin{figure}[htbp]
\includegraphics[width=\linewidth]{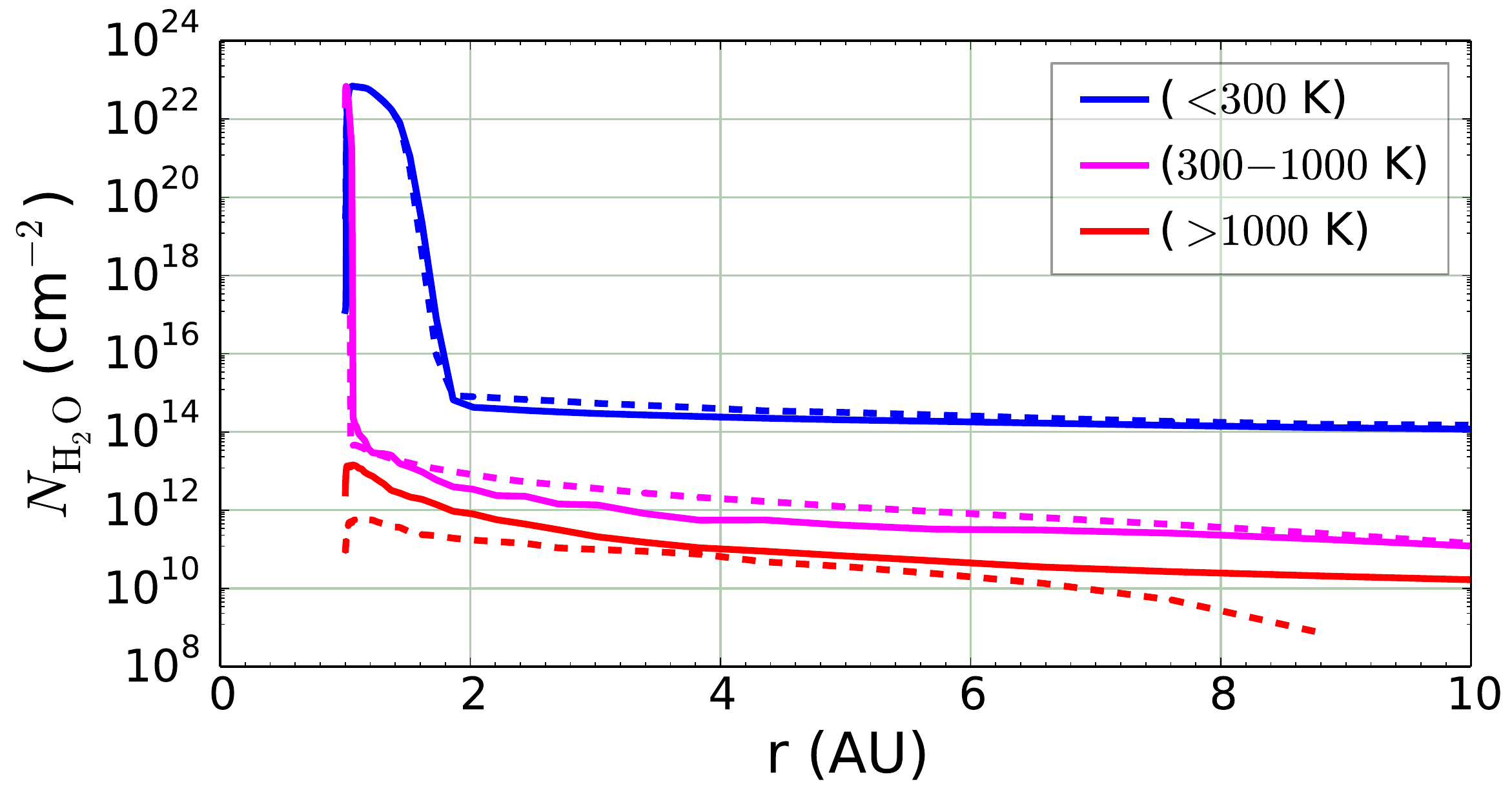}
\caption{Column density profile of water vapor in three different temperature
ranges, for a blackbody stellar spectrum ($T_\text{eff}=4000$~K, dashed lines),
and for a stellar spectrum with UV excess (solid line).
\label{figBBvsNonBB}}
\end{figure}

To see the effect of assuming the stellar spectrum to be a blackbody,
\reffig{figBBvsNonBB} shows the column density profile of water vapor in
different temperature ranges with (solid line) or without (dashed line) UV
excess in the input stellar spectrum, where the stellar temperature is 4000 K
and the UV excess is taken to be the same as in the fiducial model.  Although a
UV excess slightly reduces the column density of the $(300-1000)$~K component,
it also increases the column density of the tenuous hot ($T>1000$~K) component,
for which UV radiation is the major heating source (through photoelectric
effect, \ce{H2} photodissociation, \ce{H2} vibrational excitation, etc.).

\subsection{The Effect of Dust Settling}

As the disk evolves, dust grains coagulate and grow to larger sizes, and
gradually settle down to the midplane (and finally get assembled into forming
planets or accreted into the star), leaving less dust in the bulk, and the dust
grains that are left at higher altitudes will preferentially have a smaller
size than those close to the midplane (but may still be larger than the ISM
dust).  Our model does not yet contain a self-consistent prescription for dust
growth and settling \citep[see, e.g.,][]{Dullemond2004a,Dullemond2005}, so we
simulate the effect of dust settling in a parameterization manner in two ways.
The first is to reduce the overall dust mass of the disk, while keeping the
dust-to-gas mass ratio equal the fiducial value over the whole disk; the second
is to reduce the scale height of the larger grains, keeping a population of
small grains well-mixed with gas, while the overall dust-to-gas mass ratio
stays the same as in the fiducial model.

\begin{figure}[htbp]
\includegraphics[width=\linewidth]{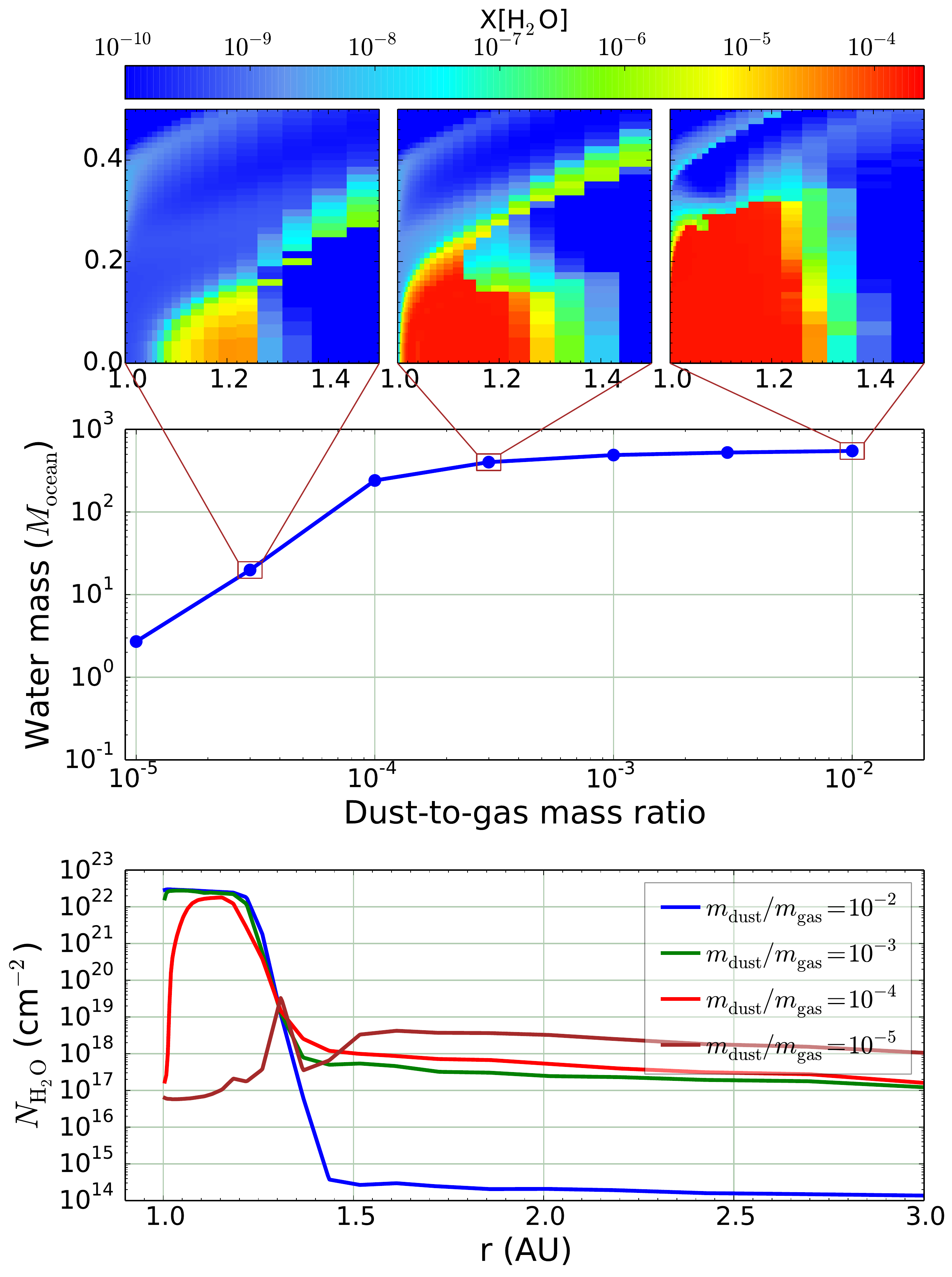}\\
\caption{Water vapor properties for different dust-to-gas mass ratios.
The top panel shows zoom-in view of water vapor distribution close to the inner
edge (1~AU).  The middle panel shows the total water vapor mass as a function
of dust-to-gas mass ratio.  The bottom panel shows the vertically integrated
water vapor column density as a function of radius for different dust-to-gas
mass ratios.
\label{figWatervsDust2Gas}}
\end{figure}

\begin{figure}[htbp]
\includegraphics[width=\linewidth]{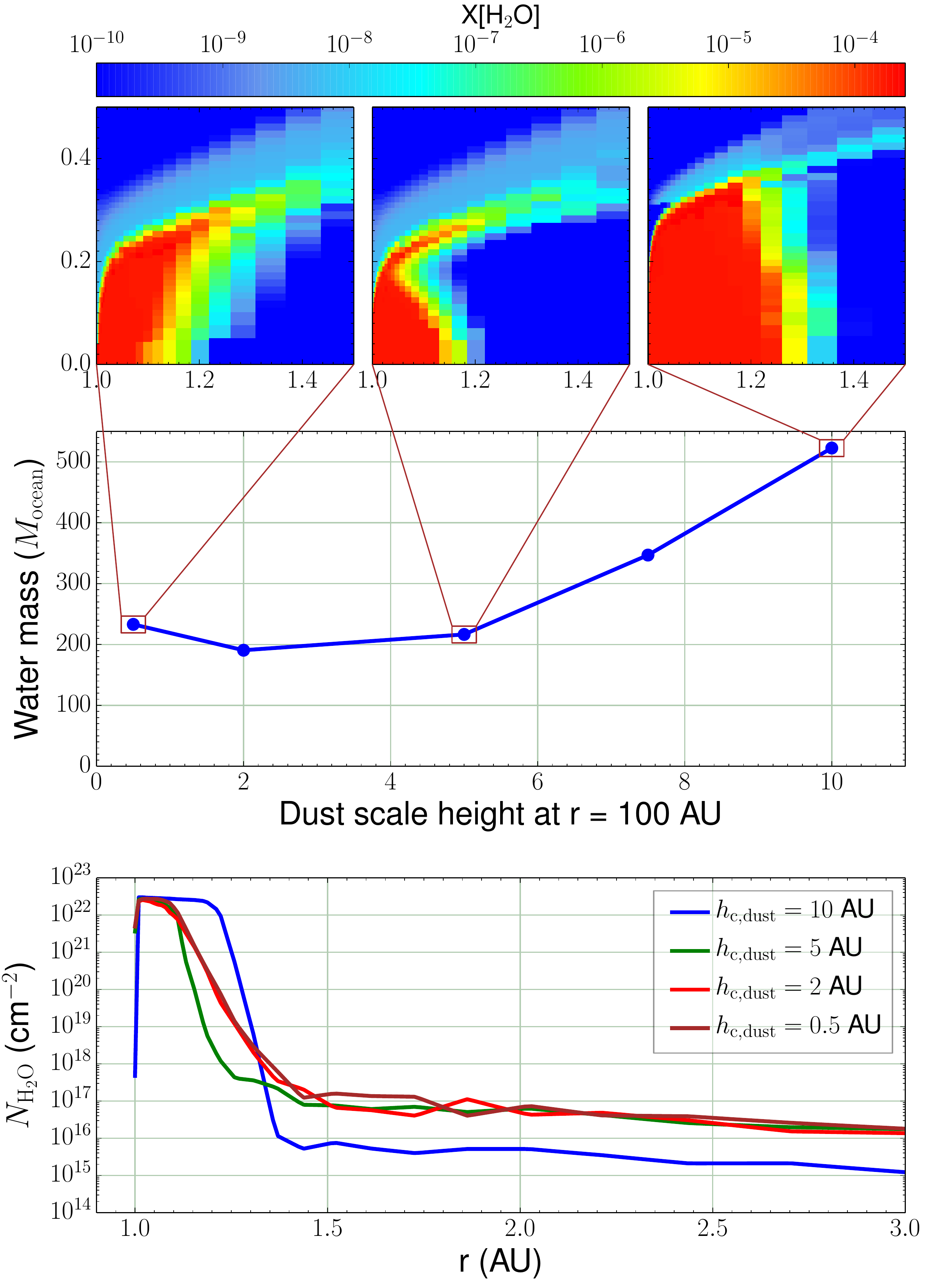}
\caption{Similar to \reffig{figWatervsDust2Gas}, except for different dust
scale heights (evaluated at $r=100$~AU) for the larger dust grains.  A
population of small dust grains are always present in these models.
\label{figWaterMassvshcDust}}
\end{figure}

The resulting warm water distribution as a function of dust-to-gas mass ratio
and of the dust scale height can be seen in \reffig{figWatervsDust2Gas} and
\reffig{figWaterMassvshcDust}.  The general trend is that less dust present in
the bulk of the disk means less warm water vapor.  The zoom-in plots in
\reffig{figWatervsDust2Gas} show that the size of the region containing
high-abundance water vapor shrinks as the amount of dust is reduced.  The main
reason is that when dust is reduced, water is more susceptible to UV
dissociation.  For example, when the dust-to-gas mass ratio is reduced from
0.01 to $10^{-4}$, the UV field strength increase by more than a factor of
$10^3$ in the region where water would otherwise be formed and preserved in gas
phase.  The self-shielding of water only starts to work at a certain depth,
depending on the density and UV intensity, but at such a depth the dust
temperature may already drops to a level lower than the water condensation
temperature.  See the next section for further discussion on this.  The
reduction of dust does have a small positive effect on water vapor abundance:
more water can be retained in the gas phase (if not photodissociated) due to
less available adsorption area.

\subsection{How Important is Water Self-shielding?}
\label{secHowImpWater}

The self-shielding of water can be calculated with
\begin{equation}
  f_{\text{sh},\ce{H2O}} = e^{-\sigma_\ce{H2O} N_\ce{H2O}},
\end{equation}
where $\sigma_\ce{H2O}$ is the photodissociation cross section of water, which
is $1.2\times10^{-17}$~cm$^{-2}$ at the frequency of Ly~$\alpha$
\citep{vanDishoeck2006}.
The dust shielding factor can be calculated with
\begin{equation}
  f_{\text{sh},\ce{dust}} = e^{-\sigma_\ce{d} N_\ce{d}}.
\end{equation}
For a silicate grain with radius of 1~$\mu$m, the absorption cross section at
$\lambda\sim0.1$~$\mu$m is about the same as the geometric cross section, hence
we may have
\begin{equation}
  \frac{\sigma_\ce{H2O} N_\ce{H2O}} {\sigma_\ce{d} N_\ce{d}} \simeq 40
  \left[\frac{X(\ce{H2O})}
  {2\times10^{-4}}\right]\left[\frac{0.01}{\eta}\right] \left[\frac{r_\text{d}}
  {1~\mu\text{m}}\right],
\end{equation}
where $\eta$ is the dust-to-gas mass ratio.
Hence the absorption due to water can be dominant over dust if water is present
in gas phase at its highest possible abundance, as shown by
\citet{Bethell2009}, which is also noted in \citet{Adamkovics2014}.

However, we have seen in the previous section that when the dust is settled or
reduced, the amount of water that is present is also reduced.  This is because
the abundance of \ce{H2} will be reduced due to the enhanced UV field and the
fact that less dust surface area is available for its formation when dust is
reduced (\ce{H2} can shield itself but nevertheless it needs dust surface to
form), which will limit the formation of \ce{H2O} from the \ce{O ->[\text{H}_2]
OH ->[\text{H}_2] H2O} chain \citep{Adamkovics2014}.  Also of importance is that even if \ce{H2} is
well shielded, CO may still be dissociated because its self-shielding is not as
efficient.  The produced C atoms can be an important competitor to the above
chain to form water.  When the gas becomes well-shielded by the dust, \ce{H2}
becomes the dominant hydrogen bearer and carbon exists as CO, and \ce{H2O} can
be formed and kept in the gas phase if the dust temperature is higher than the
water condensation temperature.  Actually, in a few test runs in which we turn
off the self-shielding of \ce{H2O} in the radiative transfer or we assume all
the oxygen not in CO is in gas phase (instead of ice) water for the initial
chemical composition, no significant changes in the resulting water vapor mass
can be seen, though some small differences are indeed noticeable (see the next
paragraph).  Hence we may say that the self-shielding of water might be
important but only in regions with rather special settings with regard to
density structure, UV intensity, and dust abundance.
Appendix~\ref{secAnaTre} contains an approximate semi-analytical account
similar to \citet{Bethell2009} for the role of water self-shielding.

\begin{figure}[htbp]
\includegraphics[width=\linewidth]{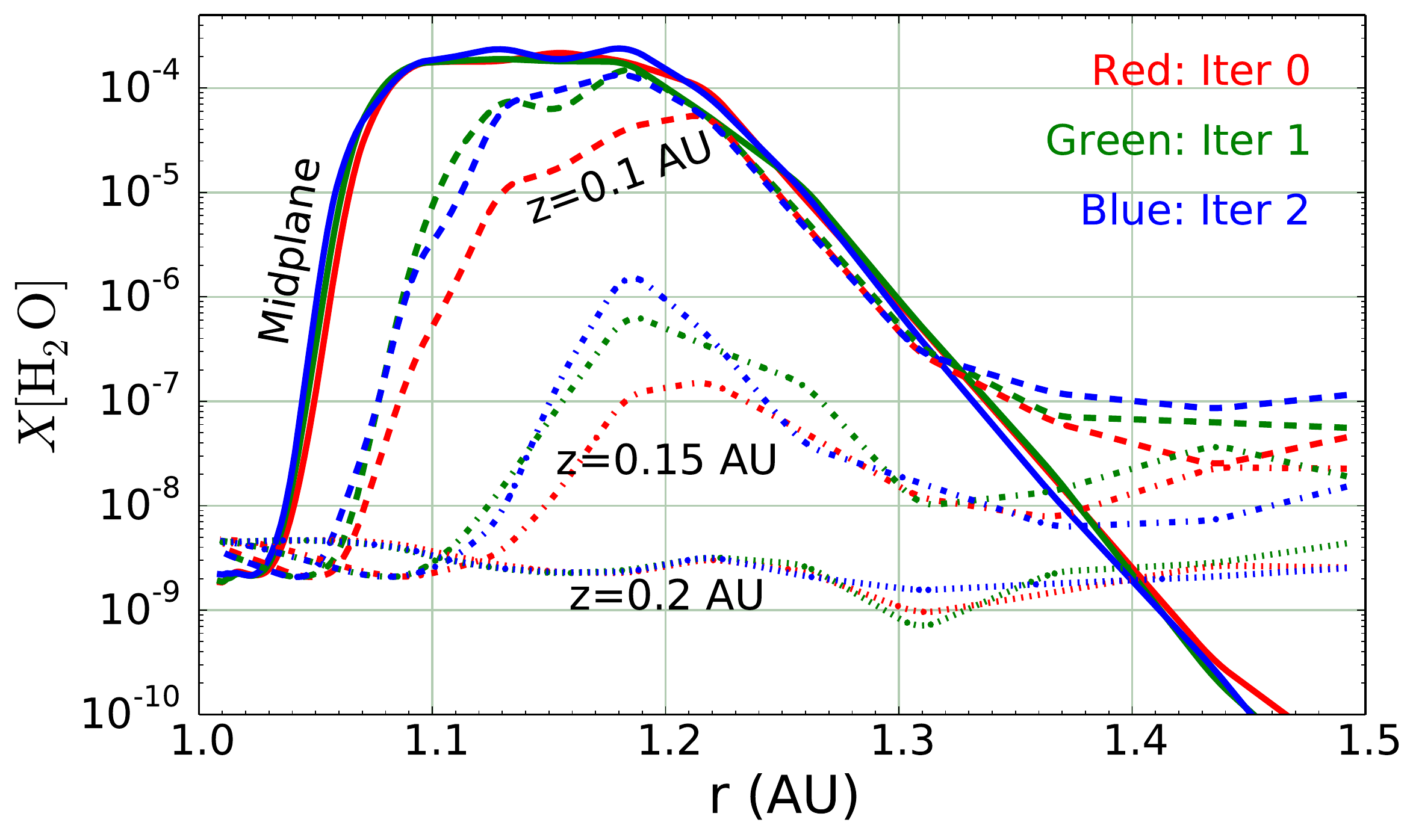}
\caption{Changes in the radial water vapor abundance profile at different
vertical height as the iteration proceeds.
\label{figWaterIteration}}
\end{figure}

The role of water self-shielding can also be checked from the output of our code.
As already noted in section~\ref{secCode}, our models work in an iterative
manner.  For iteration~0, the initial run, only dust absorption and scattering
are included, while water absorption and atomic hydrogen scattering are
\emph{not} included in the radiative transfer.  The radiative transfer outputs
the distribution of dust temperature and radiation intensity over the whole
disk, which is used in the chemical and thermal calculations.  The updated
chemical composition, specifically, the distribution of H and \ce{H2O}, are
used in the radiative transfer in the next iteration.  This process goes on
until the abundance distribution of major species (such as H, \ce{H2}, CO,
\ce{H2O}) do not vary appreciably, though the randomness inherited from the
Monte Carlo radiative transfer makes convergence in the usual sense difficult
to achieve.  \reffig{figWaterIteration} shows changes in the radial water vapor
abundance profiles at different vertical height as the iteration goes on.  The
dust-to-gas mass ratio is set to $10^{-4}$, and all the rest parameters are the
same as the fiducial model.  A close look at this figure shows that the
curves for second and third iterations extend inward towards the central star
relative to the curve of the first iteration, due to the self-shielding of
water.  For the midplane where the gas density is very high
($10^{14}$~cm$^{-3}$), the relative shift is rather small and almost
unnoticeable in the figure.  For $z=0.1$ and $0.15$~AU the effect is more
obvious, and the peak abundance of water vapor is also raised by one order of
magnitude, similar to what was found by \citet[supporting online
material]{Bethell2009}.  Higher in the disk atmosphere, with $z=0.2$~AU, where
the density drops to about $10^{13}$~cm$^{-3}$, no relative shift can be
noticed among different iterations.  As calculated semi-analytically in
Appendix~\ref{secAnaTre} (\reffig{figShieldingDepth}), with a density of
$10^{13}$~cm$^{-3}$, a dust-to-gas mass ratio of $10^{-4}$, and a strong UV
field, the depth for water to be shielded is $\sim$0.5~AU.  But at depth 0.5~AU
into the disk the dust temperature will be just low enough (see
\reffig{figGasTemperature}, or maybe more clearly, \reffig{figTAnafitting}) for
water to condense out, and then the water vapor abundance will be determined by
photo-desorption.

\section{Discussion and Conclusions}
\label{secConclusion}

We have modeled the formation of warm water vapor in protoplanetary disks with
a comprehensive model.  The radiative transfer of UV continuum and
Ly~$\alpha$ photons and the associated heating of dust grains are calculated
with a Monte Carlo method.  The density structure is described in a
parameterized manner, and the gas temperature structure is solved based on the
balance between heating and cooling mechanisms.  The chemical evolution is
followed for 1~Myr.

We find that warm water is mainly distributed in a small region close to the
inner edge of the disk.  The location and size of this region is determined by
two factors: the attenuation of the dissociating UV photons by dust grains and
water molecules, and the condensation of water onto dust grains when the dust
temperature abruptly drops below $\sim$150~K.  At high densities and if dust is
not severely settled, the attenuation length is very short (see
Appendix~\ref{secAnaTre}), so water vapor can exist right next to the edge
\citep{Cleeves2011}.  The diffusion and escape of photons creates a steep
temperature drop at the inner edge, which limits the size of the region in
which warm water vapor resides.  Assuming the same disk structure, a more
luminous central star leads to a wider region where gas phase water is found in
abundance.  As dust grows and gets settled to the midplane, the region
containing warm water vapor also shrinks, though a significant amount of water
vapor can still exist if a population of small dust grains remain in the bulk
of the disk.

Observationally, the concentration of warm water vapor in the inner disk will
produce a sharp ring structure, which is, in general, in agreement with
analysis of water vapor emission.  We do note that, specifically in the case of
TW~Hya, the column density of water vapor at $\sim$4~AU predicted by our model
is lower than the best-fit value of \citet{Zhang2013} by three orders of
magnitude, unless we use a desorption energy of water lower than what is
experimentally determined.  Also for
TW~Hya, the amount of diffuse cold water vapor produced by photodesorption in
our model is close to the observed value of \citet{Hogerheijde2011}.
Whether the discrepancy (and agreement) here should be taken seriously
can only be answered with detailed radiative transfer modeling based on the
chemical structure, which will be the topic of a future paper.  

It might be possible to use warm water vapor as a tracer of the
kinetics of the inner disk, since it is concentrated in a small inner region
\citep{Pontoppidan2010b} at or close to where planets form.  In the era of
ALMA, it may also become possible to directly map the distribution of spectrally
resolved lines of water vapor (or rather its isotopologue \ce{H2$^{18}$O} to
avoid absorption from the telluric water line) in the inner region of nearby
protoplanetary disks.  
For example, the $4_{14}$ -- $3_{21}$ line of
\ce{H2$^{18}$O} at 390.60776~GHz will become optically thick if
$N(\ce{H}_2^{16}\ce{O}) \gtrsim 2{\times}10^{19}$~cm$^{-2}$, assuming a
$^{16}$O/$^{18}$O ratio of 500 and $T\gtrsim 300$~K.  With the highest possible
resolution of the full ALMA array at this wavelength (${\sim}0.01''$), a warm
water vapor ring with radius $\sim$1~AU in a protoplanetary disk 100~pc away is
marginally resolvable.

In Appendix~\ref{secAnaTre} we semi-analytically study the importance of the
shielding due to water and OH.  They can be important in some part of the
parameter space (high density, low but nonzero dust-to-gas mass ratio,
intermediate UV field), but their role is likely to be too subtle to be
of paramount importance in generic settings.

\acknowledgments We thank N.~Calvet and L.~Hartmann for useful discussions on
radiative transfer and viscous heating.  This work is supported by grant
NNX12A193G from the NASA Origin of Solar Systems Program.

\appendix

\twocolumn

\section{Radial Temperature Profile Close to the Disk Inner Edge}
\label{secAnaTProfile}

\begin{figure}[htbp]
\includegraphics[width=\linewidth]{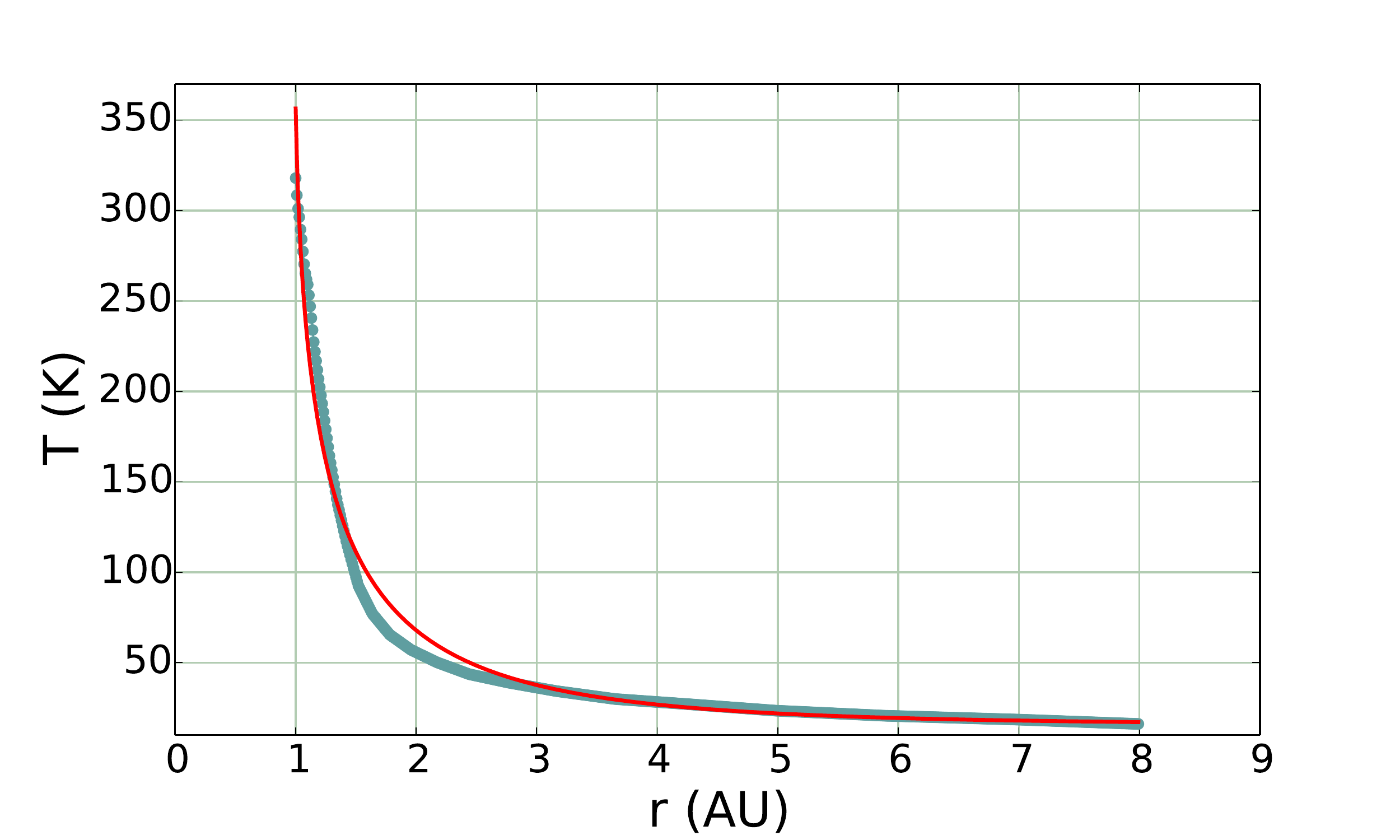}
\caption[]{Mid-plane temperature of the inner disk as a function of radius,
overlapped with an analytical fitting (red thin line).
\label{figTAnafitting}}
\end{figure}

When the medium is very opaque, the transfer of radiation can be approximated
by a diffusion process, which gives the following equation for the temperature
gradient
\begin{equation}
  \frac{\partial T}{\partial r} = -\frac{3}{64\pi \sigma_\text{SB}}
  \frac{\kappa_\text{R} \rho l f}{r^2 T^3}.
\end{equation}
This equation is a modified version of the one in \citet{Kippenhahn2013}.  In
the above equation, $\sigma_\text{SB}$ is the Stefan-Boltzmann constant,
$\kappa_\text{R}$ is the Rosseland mean opacity, $\rho$ is the density, $l$ is
the stellar luminosity, and $f$ is a function of $r$ to account for the
``leakage'' of radiation from the upper and lower surfaces of the disk.
Without the $f$
factor, the equation only applies for spherical geometry.  At low
temperature, $\kappa_\text{R}$ is roughly proportional to $T^2$
\citep{Kruegel2008}.  For a disk model with $\Sigma(r)\propto r^{-1}$ and
$h\propto r$, we have
$\rho\propto r^{-2}$.
For the function $f$, we may qualitatively parameterize it as
\begin{equation}
  f(r) = \frac{a r^2}{(a+1) r^2 - r^2_\text{in}},
\end{equation}
where $r_\text{in}$ is the disk inner radius, and the empirical parameter
$a\ll1$ is roughly the fraction of energy transported radially outward through
the disk edge (rather than through the upper and lower surface).  We thus have
\begin{equation}
  \frac{\partial T}{\partial r} \propto -T^{-1} r^{-4} f(r).
  \label{eqdTdr}
\end{equation}
Here we are only concerned with the inner region (and the approximation we use
here does not apply to the outer region anyway), so we may assume
\begin{equation*}
x\equiv(r-r_\text{in})/r_\text{in}\ll1.
\end{equation*} 
An approximate solution in this regime is
\begin{equation}
  T = T_\text{in} \left[1 - \frac{3a}{16}\frac{r_\text{in}}{l_\text{PH}}
  \ln\left(1 + \frac{2}{a}x\right)\right]^{1/2},
\end{equation} 
where $T_\text{in}$ is the dust temperature right at the inner edge,
and $l_\text{PH}\equiv (\kappa_\text{R}\rho)^{-1}$ is the mean free path of
photons (emitted by dust of temperature ${\sim}T_\text{in}$).
For $T_\text{in}=300$~K and $n=10^{14}$~cm$^{-3}$, using the calculated
Rosseland opacity from \citet{Semenov2003}, we have $l_\text{PH}\sim
0.01$~AU.  Take $r_\text{in}=1$~AU and $a=0.01$, we then have
\begin{equation*}
  T = T_\text{in} \left[1 - 0.2\ln\left(1+200x\right)\right]^{1/2},
\end{equation*} 
which means that an outward shift of only 0.2~AU from the inner edge can reduce
the temperature by half.
A fitting based on the analytical integration of \refeq{eqdTdr} to the
midplane dust temperature profile (calculated from Monte Carlo simulation) of
the inner disk is shown in \reffig{figTAnafitting}, which is not perfect but
captures the major trend.


\section{An Analytical Treatment for Warm Water Formation under
Photodissociation}
\label{secAnaTre}

\begin{figure}[htbp]
\centering
\includegraphics[width=\linewidth]{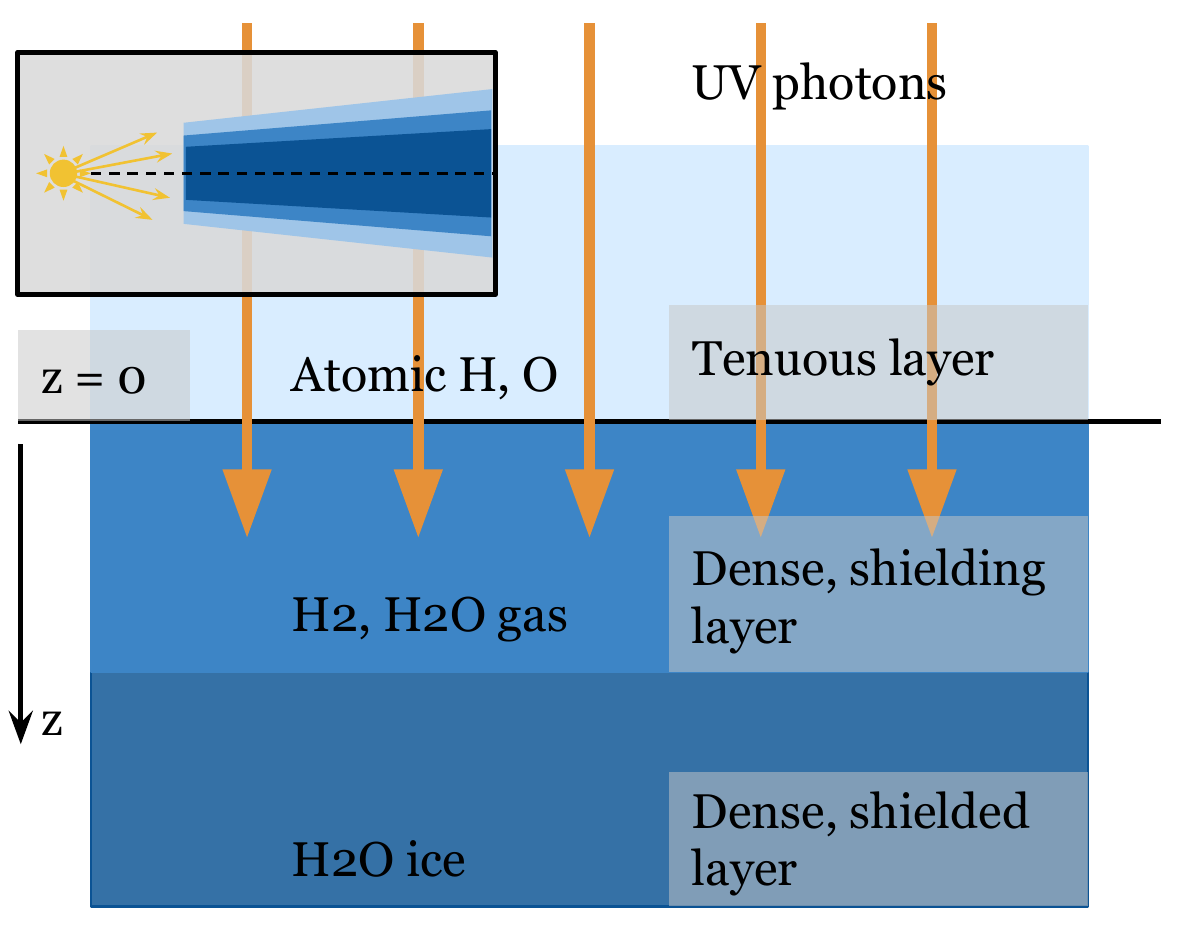}
\caption{Sketch of the analytical toy model.  Here we are mainly concerned with
the tenuous layer and the shielding layer.  The small inset shows the context
where this toy mode may be applied.}
\label{figToyGeometry}
\end{figure}

Here we present an approximate analytical treatment of water formation with
warm neutral chemistry under the effect of UV photo\-dissociation.  This
treatment is very similar to the one in \citet{Bethell2009}, except that here
we include the formation and photo\-dissociation of \ce{H2}, and the geometry
we assume is much simpler.  As sketched in \reffig{figToyGeometry}, we consider
a uniform and isothermal slab of gas and dust irradiated by external UV field.
To put in context, this slab may be viewed as one horizontal slice of the inner
disk, and the irradiation comes from the central star.
The goal here is \emph{not} to imitate a realistic circumstellar disk, but
rather to see to what extent is water (and OH) shielding important.

We consider water formation through hot neutral chemistry and destruction
by UV photons:
\begin{align}
  \ce{H2 + O } & \ce{-> OH + H}, \label{eqOHf1} \\
  \ce{H2 + OH} & \ce{-> H2O + H}, \label{eqH2Of1} \\
  \ce{H2O} & \ce{->[\text{UV}] OH + H}, \label{eqH2Ophd1} \\
  \ce{OH}  & \ce{->[\text{UV}] O  + H}. \label{eqOHphd1}
\end{align}
For simplicity competitive channels such as \ce{C + OH ->
CO + H} and \ce{O + OH -> O2 + H} are omitted.  They can be important in
regions where CO and \ce{H2} are partially dissociated or where the temperature
is low ($\lesssim$200~K).
The four reactions gives the following set of equations:
\begin{align}
  \partial_t n(\ce{H2O}) &= -\kiii n(\ce{H2O}) + \kii n(\ce{H2}) n(\ce{OH}), \\
  \partial_t n(\ce{OH})  &= -\kiv n(\ce{OH}) - \kii n(\ce{H2}) n(\ce{OH}), \\
                         &\quad  + \ki n(\ce{H2}) n(\ce{O}) + \kiii n(\ce{H2O}), \\
  \partial_t n(\ce{O}) &= -\ki n(\ce{H2}) n(\ce{O}) + \kiv n(\ce{OH}).
\end{align}
We further assume that oxygen is completely contained in the three species:
\begin{align}
  n(\ce{H2O}) + n(\ce{OH}) + n(\ce{O}) = n_\ce{O},
\end{align}
where $n_\ce{O}$ is the density of oxygen nucleus.

Assuming steady-state condition, we can readily solve the above equations to
get
\begin{align}
  n(\ce{OH}) &= n(\ce{H2O}) \frac{\kiii}{\kii n(\ce{H2})}, \label{eqOH_H2O} \\
  n(\ce{O}) &= n(\ce{OH}) \frac{\kiv}{\ki n(\ce{H2})}, \\
  n(\ce{H2O}) &= \frac{n_\ce{O}}
                {\displaystyle 1 + \frac{\kiii}{\kii n(\ce{H2})}
                 \left(1 + \frac{\kiv}{\ki n(\ce{H2})}\right) }. \label{eqH2Odens}
\end{align}
The above three relations hold at each point of the slab.  Since the parameters
$n(\ce{H2})$, $\kiii$, and $\kiv$ in the right hand side may change with depth,
so will the variables in the left hand side, which is what we will solve in
the following.

At the present we are mainly concerned with the initial growth of $n(\ce{H2O})$
as a function of $z$, so we can assume
\begin{equation*}
  \frac{\kiv}{\ki n(\ce{H2})} \gtrsim \frac{\kiii}{\kii n(\ce{H2})} \gg 1.
\end{equation*} 
If this assumption does not hold, then we would already have $n(\ce{H2O})\sim
n_\ce{O}$, and no further discussions are needed.  Hence we may first
approximate \refeq{eqH2Odens} by
\begin{align}
  n(\ce{H2O}) &= n_\ce{O} \frac{\ki \kii n^2(\ce{H2})} {\kiii \kiv}
  \tag{\ref{eqH2Odens}$'$} \label{eqH2Odens1}
\end{align}
This will slightly \emph{overestimate} the abundance of \ce{H2O}.

The photodissociation rate $\kiii$ and $\kiv$ can be written as
\begin{equation}
\begin{split}
  \kiii &= \kiiiz \exp\left[-\sigiii N(\ce{H2O}) - \sigiv N(\ce{OH})
                           -\sigma_\ce{d}N_\ce{d}\right], \\
  \kiv  &= \kivz \exp\left[-\sigiii N(\ce{H2O}) - \sigiv N(\ce{OH})
                           -\sigma_\ce{d}N_\ce{d}\right],
\end{split}
\label{eqkiii}
\end{equation}
in which the self-shielding of water and \ce{OH} and the shielding due to dust
are included.  $\kiiiz$ and $\kivz$ are the unshielded rate, $\sigiii$ and
$\sigiv$ are the dissociation cross section of water and \ce{OH}, and
$\sigma_\ce{d}$ is the dust absorption cross section.  $N(\ce{H2O})$,
$N(\ce{OH})$, and $N_\ce{d}$ are the respective column densities.

Combining \refeq{eqH2Odens1} and \refeq{eqkiii}, we have
\begin{equation}
\begin{split}
  & \sigiii N(\ce{H2O}) + \sigiv N(\ce{OH}) + \sigma_\ce{d}N_\ce{d}  \\
  =& \frac{1}{2}\ln\left(\frac{n(\ce{H2O})\kiiiz \kivz}{n_\ce{O} \ki \kii
  n^2(\ce{H2})} \right)
\end{split}
\label{eqBeforeDiff}
\end{equation}
Differentiating both sides with respect to $z$ gives
\begin{equation}
\begin{split}
  &\quad \frac{1}{n(\ce{H2O})} \ddz n(\ce{H2O}) -
         \frac{2}{n(\ce{H2})} \ddz n(\ce{H2}) \\
  &= 2(\sigiii n(\ce{H2O}) + \sigiv n(\ce{OH}) + \sigma_\ce{d} n_\ce{d}) \\
  &= 2\Big[ \sigiii n(\ce{H2O}) \\
  & \qquad + \sigiv \left(\frac{\kiiiz \ki n_\ce{O}} {\kivz \kii
    n(\ce{H2O})}\right)^{1/2} n(\ce{H2O}) \\
  & \qquad + \sigma_\ce{d} n_\ce{d} \Big],
\end{split}
\end{equation}
where we have used \refeq{eqOH_H2O}, (\ref{eqkiii}), and (\ref{eqBeforeDiff}),
and have assumed temperature and density are constant.

Define $x \equiv n(\ce{H2O}) / n_\ce{O}$, $x_\ce{d} \equiv n_\ce{d} /
n_\ce{O}$, we get
\begin{equation}
\begin{split}
  \ddz x &= 2n_\ce{O} \left[\sigiii x^2 + \sigiv \left(\frac{\kiiiz \ki}{\kivz
  \kii}\right)^{1/2} x^{3/2} \right. + \\
  &\qquad\qquad \sigma_\ce{d} x_\ce{d} x
   \left.\vphantom{\left(\frac{\kiiiz \ki}{\kivz\kii}\right)^{1/2}}\right]
  + \frac{2x}{n(\ce{H2})} \ddz n(\ce{H2}).
\end{split}
\label{eqDxDz}
\end{equation}
This equation only works for $x\ll1$.


The \ce{H2} abundance can be calculated by
\begin{equation}
\begin{split} \nspe{H2} =  n_\ce{H} \times
 \frac{n_\ce{d}\tilde{\sigma}_\ce{d}v_\text{T}/2} {\xi
  + n_\ce{d}\tilde{\sigma}_\ce{d}v_\ce{T}},
\end{split}
\label{eqnH2}
\end{equation}
where $\tilde{\sigma}_\ce{d}$ is the dust particle cross section for adsorbing
\ce{H} atoms (to be distinguished from the cross section $\sigma_\ce{d}$ for
absorbing UV photons), $v_\text{T}$ is the average velocity of \ce{H} atoms,
and $\xi$ is the total dissociation (photo + cosmic-ray) rate of \ce{H2}.
In calculating $\xi$ we take into account the dust attenuation and the
\ce{H2} self-shielding.  Since the self-shielding of \ce{H2} involves the column
density of \ce{H2} \citep{Draine1996}, we first treat \refeq{eqnH2} as a
differential equation of $N(\ce{H2})$ since
$\mathrm{d}N(\ce{H2})/\mathrm{d}z=n(\ce{H2})$, and after $N(\ce{H2})$ at depth
$z$ is solved, its value is feed back to \refeq{eqnH2} to get $n(\ce{H2})$ at
$z$.  This is similar to \citet[p289]{Tielens2005}.

When solving \refeq{eqDxDz} it is important to get the boundary value $x|_{z=0}$
right.  We calculate this value based on \refeq{eqH2Odens}, but corrected for
the competitive reactions involving O and C, though these reactions are not
included in the integration.  In the calculation the dust property is taken to
be the same as in the fiducial model (Section \ref{secFiducial}).
We adopt the following rate parameters:
\begin{equation}
\begin{split}
  \ki  &= 3.14\times10^{-13} \left(\frac{T}{300}\right)^{2.7}
          e^{-3150/T}~\text{cm}^3~\text{s}^{-1}, \\
  \kii &= 2.05\times10^{-12} \left(\frac{T}{300}\right)^{1.52}
          e^{-1736/T}~\text{cm}^3~\text{s}^{-1}, \\
  \kiiiz &= 1.2\times10^{-4}
    \left(\frac{\FLya}{10^{13}~\text{cm}^{-2}~\text{s}^{-1}}\right)
    ~\text{s}^{-1}, \\
  \kivz &= 1.8\times10^{-5}
    \left(\frac{\FLya}{10^{13}~\text{cm}^{-2}~\text{s}^{-1}}\right)
    ~\text{s}^{-1}, \\
  \sigiii &= 1.2\times10^{-17}~\text{cm}^{2}, \\
  \sigiv &= 1.8\times10^{-18}~\text{cm}^{2},
\end{split}
\end{equation}
where $\FLya$ is the number flux of Ly~$\alpha$ photons.
We assume $T=300$~K.

\begin{figure}[htbp]
\centering
\includegraphics[width=\linewidth]{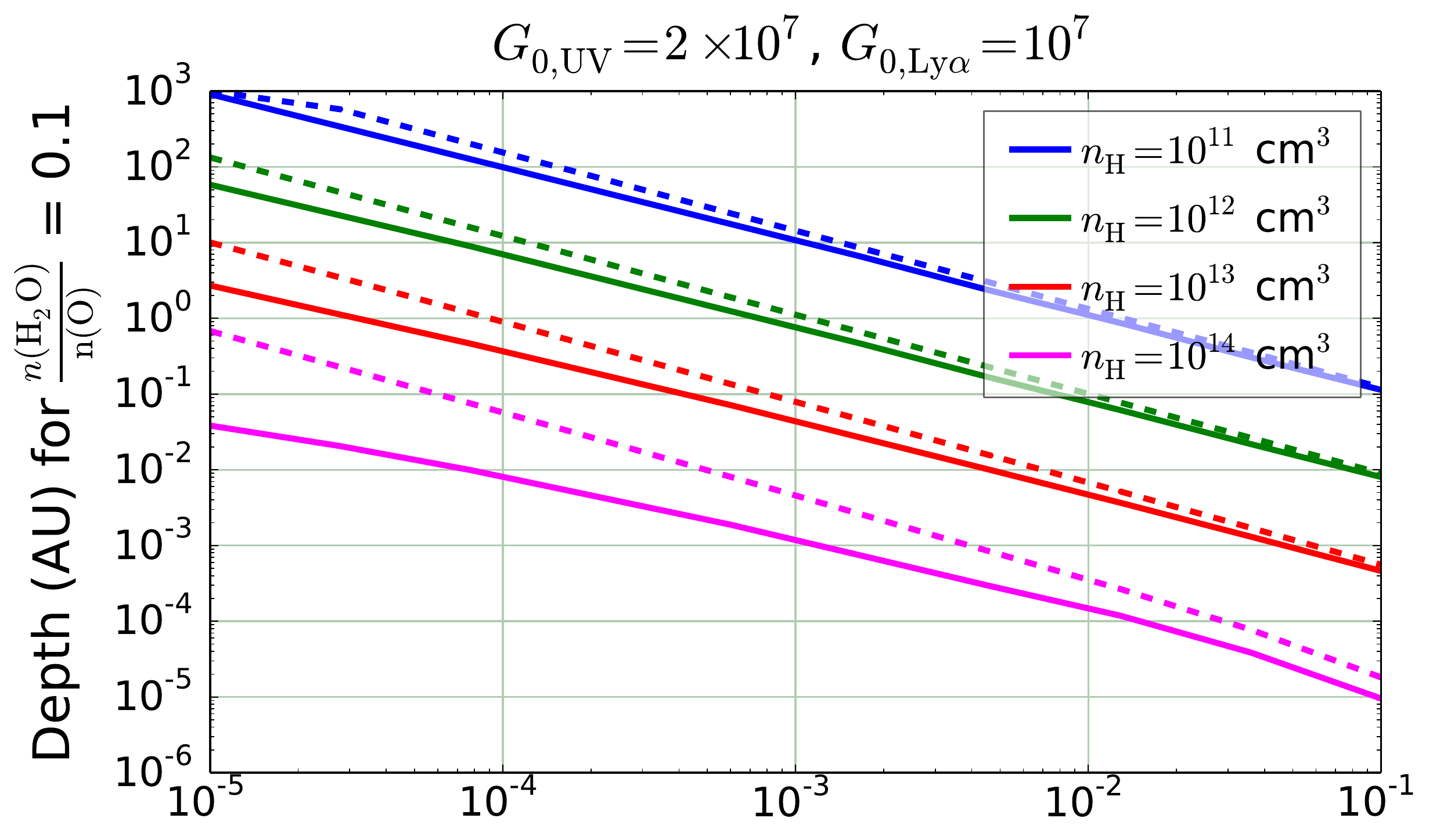}
\includegraphics[width=\linewidth]{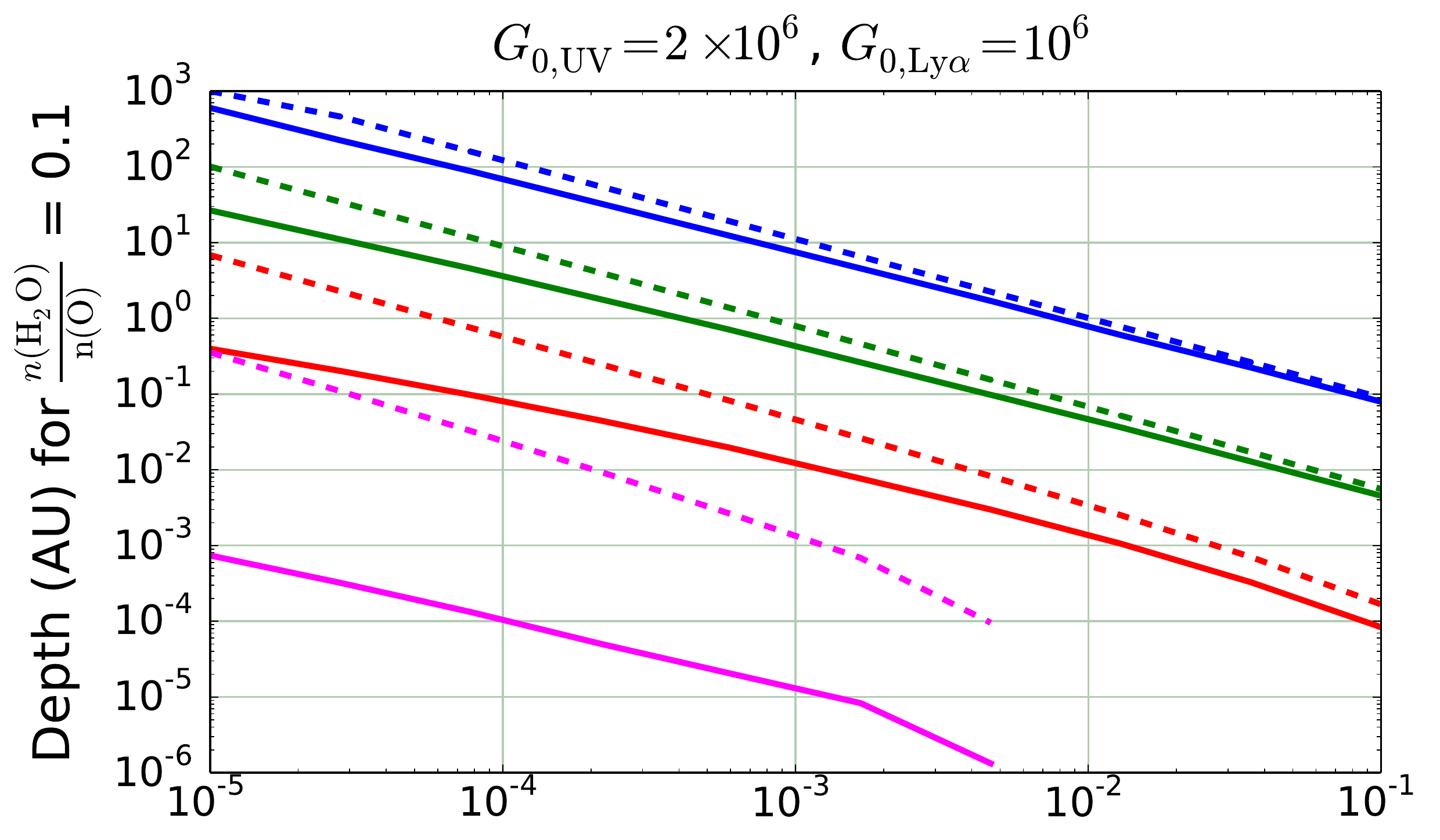}
\includegraphics[width=\linewidth]{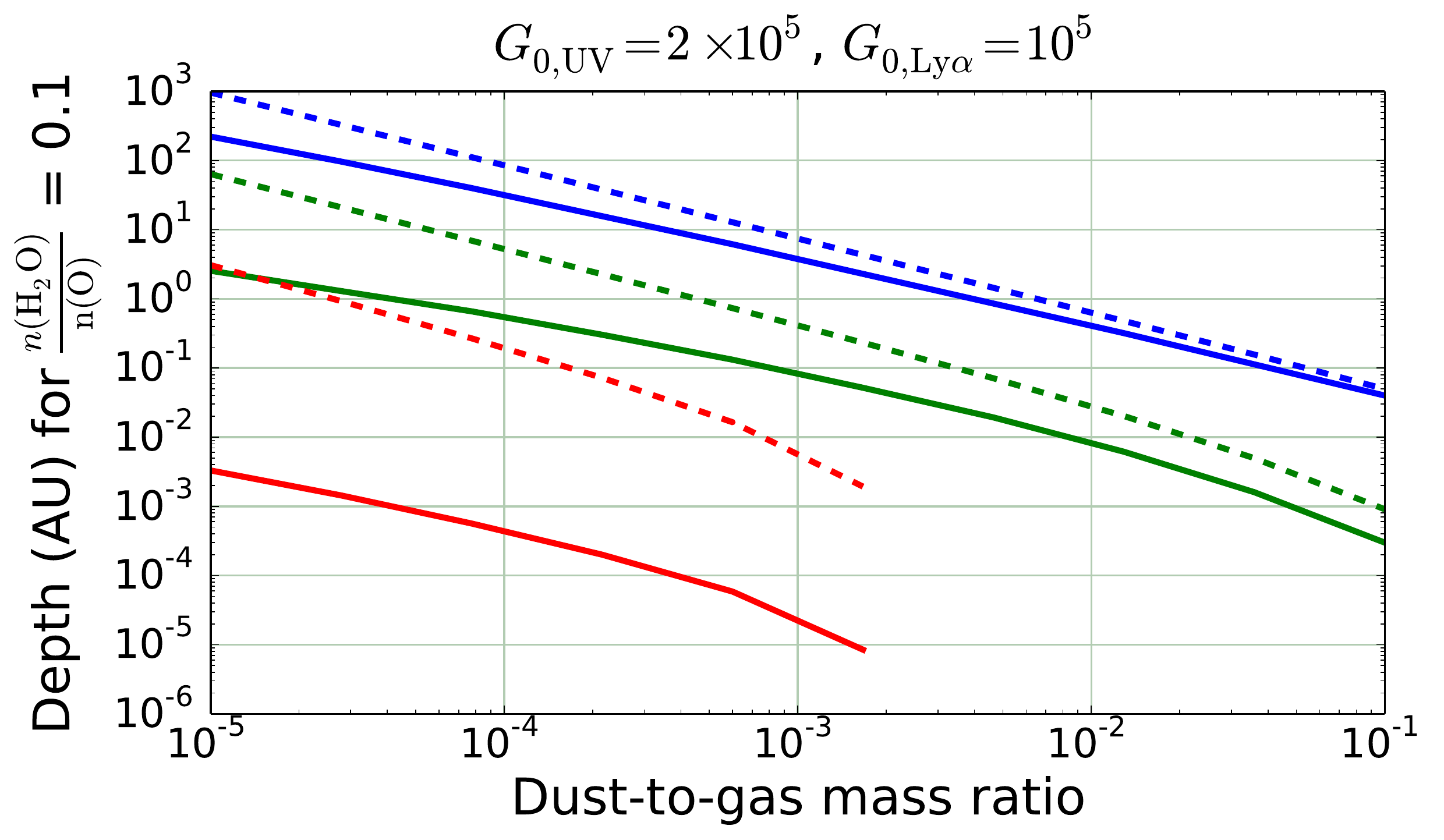}
\caption{Shielding depth of \ce{H2O} (the depth at which $n(\ce{H2O})$
reaches $0.1 n_\ce{O}$) as a function of dust-to-gas mass
ratio for different gas densities.  The top panel has the strongest UV fields, and the bottom one has the weakest.
The solid lines are for cases in which the shielding due
to dust, \ce{H2O}, and OH are all considered, while the dashed
lines are for cases in which only dust shielding is included.
The magenta curves (for the $n_\ce{H}=10^{14}$~cm$^{-3}$ case) are below
the lower plotting range in the bottom panel.
\label{figShieldingDepth}}
\end{figure}

The importance of the shielding due to \ce{H2O} and OH can be tested by
considering the changes in the depth (we call it ``shielding depth'') at which
water reaches a ``high'' abundance, which we arbitrarily take to be 10\% of
the total available oxygen abundance.  The shielding depth as a function of the
dust-to-gas mass ratio is plotted in \reffig{figShieldingDepth} for different
densities.  The solid lines are for cases which include the shielding due to
dust, \ce{H2O}, and OH, while the dashed lines are for cases in which the
shielding due to \ce{H2O} and OH are turned off.  Different panels have a
different UV continuum and Ly~$\alpha$ intensity.

\begin{figure}[htbp]
\centering
\includegraphics[width=\linewidth]{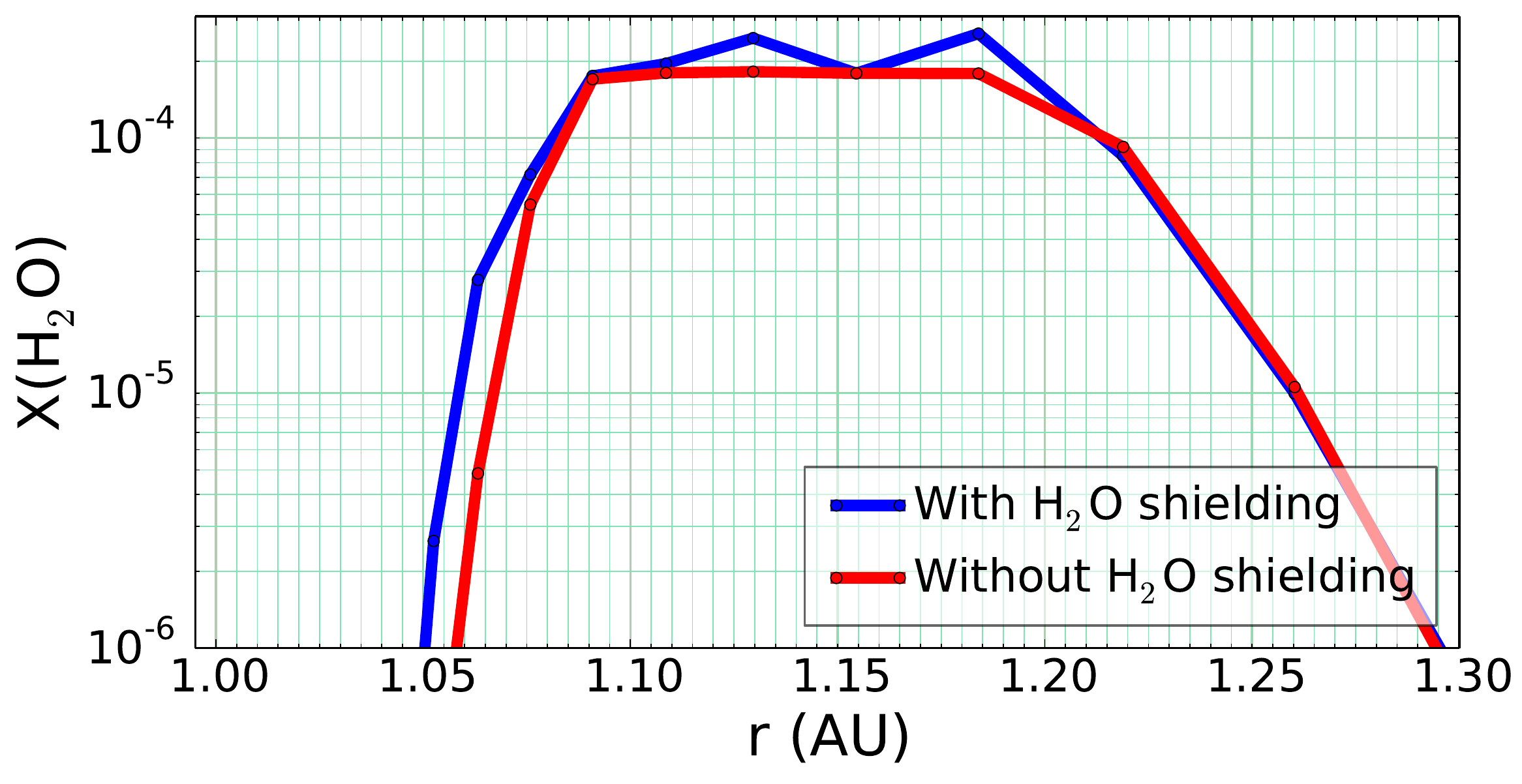}
\caption{Midplane water abundance as a function of radius for two cases
with or without the shielding due to \ce{H2O} included.  The dust-to-gas mass
ratio is $10^{-4}$.  The disk inner edge is at 1~AU.  Note that they are
calculated from our full code, not from the simple semi-analytical model
presented in this appendix, hence the shielding depths are not as shown in
\reffig{figShieldingDepth}.}
\label{figDiffH2O}
\end{figure}

As can be seen in \reffig{figShieldingDepth}, for lower densities
($n_\text{H}=10^{11}$~cm$^{-3}$) and strong UV fields, the shielding depth is
rather large ($\gtrsim$1~AU) for normal or small dust-to-gas mass ratios, which
indicates that in reality water cannot form through the hot neutral chemistry
at such densities (if the UV radiation is as strong as we have assumed here),
because at the calculated shielding depth the dust temperature might have
decreased to the condensation temperature of water.  Also can be seen is that
the effect of turning-off the shielding due to water and OH is not significant
for low densities, high dust-to-gas mass ratio, or strong UV fields, but
becomes significant otherwise.  For example, for
$n_\text{H}=10^{14}$~cm$^{-3}$, in the strong UV field case, when the
dust-to-gas mass ratio is reduced to $0.1$ -- $10^{-3}$ times the normal ISM
value, turning off \ce{H2O} and OH shielding increases the shielding depth by a
factor of a few to ten, while in the weak UV field case the shielding depth can
increase by more than two orders of magnitude.  On the other hand, the
shielding depths at such high densities are small anyway, hence the changes may
not have obvious practical effects.  For example, \reffig{figDiffH2O} shows the
radial water abundance profile at the midplane
($n_\text{H} \sim 10^{14}$~cm$^{-3}$) for dust-to-gas mass ratio equal
$10^{-4}$, with or without water shielding.  Without water shielding the
profile recedes slightly further away from the inner edge.  With even smaller
dust-to-gas mass ratio the distinction will be larger.  However, as described
above for the low density case, since the shielding depth increases as
dust-to-gas mass ratio decreases, at some point the shielding depth will be
large enough that the dust temperature drops below the water condensation
temperature, and it becomes impossible to keep abundant water in the gas
phase.  In the extreme case, when there is no dust, then the problem becomes
that \ce{H2} cannot be maintained (for $G_0=10^7$ the photodissociation time
scale of \ce{H2} is of the order of days) for water to form in the first place.
Another issue is that, when the dust becomes optically thin, the dust
temperature may also drop because less radiation can be trapped, which makes
water more likely to condense out unless the total surface area of dust grains
has been extremely reduced.

In the above discussions we did not include chemical reactions involving
excited \ce{H2} molecules.  UV radiation can excite \ce{H2} molecules to
vibration levels with $v>0$, which is capable of increasing the rate
coefficients profoundly \citep{Agundez2010}.  However, the net effect depends
on the abundance of excited \ce{H2}, which depends on the UV intensity and
collisional deexcitation rates.  For density higher than
${\sim}10^{11}$~cm$^{-2}$ and $G_0\sim10^7$, we estimate that the abundance of
excited \ce{H2} will be too low to affect the discussion here.

As a side note regarding \reffig{figDiffH2O}, the fact that the depth at which
the water abundance reaches higher than 10\% of the total oxygen abundance is
larger than what would be expected from the top panel of
\reffig{figShieldingDepth} (0.05~AU versus 0.01~AU) is due to the activation of
hot neutral chemistry that destroys water.  Since the dust-to-gas mass ratio is
much reduced, the accommodation cooling is not effective and the gas
temperature can be much higher than the dust temperature ($\sim$1500~K versus
$\sim$200~K).  At this temperature not only \ce{H2O} gets destroyed by reacting
with H atoms, \ce{H2} is also reduced to a low abundance ($\sim$0.01) due to
reaction with OH.  A simple analytical model cannot easily capture all these
features.



\clearpage

\bibliographystyle{apj}

\end{document}